\documentclass[preprint,12pt]{elsarticle}
\usepackage[top=1in, bottom=1in, left=1in, right=1in]{geometry}
\usepackage{rotating}

\usepackage{graphicx}
\usepackage{amssymb}

\usepackage{lineno}
\usepackage{comment}
\usepackage{amsmath}
\usepackage{tabu}
\usepackage{xcolor}
\usepackage{setspace}
\usepackage{bm}
\usepackage[colorlinks=true,bookmarksopen,pdfsubject={algorithms},linkcolor={blue}, anchorcolor={black}, citecolor={blue}, filecolor={magenta}, menucolor={black}, pagecolor={red},backref=none,urlcolor={blue}]{hyperref}
\usepackage[ruled,vlined]{algorithm2e}
\usepackage{listings}
\usepackage{xcolor}

\definecolor{codegreen}{rgb}{0,0.6,0}
\definecolor{codegray}{rgb}{0.5,0.5,0.5}
\definecolor{codepurple}{rgb}{0.58,0,0.82}
\definecolor{backcolour}{rgb}{0.95,0.95,0.92}

\lstdefinestyle{mystyle}{
  backgroundcolor=\color{backcolour}, commentstyle=\color{codegreen},
  keywordstyle=\color{magenta},
  numberstyle=\tiny\color{codegray},
  stringstyle=\color{codepurple},
  basicstyle=\ttfamily\footnotesize,
  breakatwhitespace=false,         
  breaklines=true,                 
  captionpos=b,                    
  keepspaces=true,                                    
  numbersep=5pt,                  
  showspaces=false,                
  showstringspaces=false,
  showtabs=false,                  
  tabsize=2
}

\journal{NPJ Computational Materials (Accepted)}

\begin{document}
\begin{spacing}{1.25}
\begin{frontmatter}

\title{Revealing Nanostructures in High-Entropy Alloys via Machine-Learning Accelerated Scalable Monte Carlo Simulation}

\author{Xianglin Liu\corref{cor1}$^\dagger$\footnote{\footnotesize{ $^\dagger$ These two authors contributed equally to this work}}$^\ast$\footnote{\footnotesize{ $^\ast$ corresponding author}}  } 
\ead{xianglinliu01@gmail.com}
\address{Pengcheng Laboratory, Shenzhen, China}

\author{Kai Yang$^\dagger$}
\address{Pengcheng Laboratory, Shenzhen, China}

\author{Yongxiang Liu}
\address{Pengcheng Laboratory, Shenzhen, China}

\author{Fanli Zhou}
\address{School of Computer and Artificial Intelligence, Xiangnan University, Chenzhou, China}

\author{Dengdong Fan}
\address{Pengcheng Laboratory, Shenzhen, China}

\author{Zongrui Pei}
\address{New York University, NY 10012, USA}

\author{Pengxiang Xu\corref{cor2}$^\ast$}
\ead{xupx@pcl.ac.cn}
\address{Pengcheng Laboratory, Shenzhen, China}

\author{Yonghong Tian}
\address{School of AI for Science, Shenzhen Graduate School, Peking University, Shenzhen, China}
\address{Pengcheng Laboratory, Shenzhen, China}

\begin{abstract}
First-principles Monte Carlo (MC) simulations at finite temperatures are computationally prohibitive for large systems due to the high cost of quantum calculations and poor parallelizability of sequential Markov chains in MC algorithms. We introduce scalable Monte Carlo at eXtreme (SMC-X), a generalized checkerboard algorithm designed to accelerate MC simulation with arbitrary short-range interactions, including machine learning potentials, on modern accelerator hardware. The GPU implementation, SMC-GPU, harnesses massive parallelism to enable billion-atom simulations when combined with machine-learning surrogates of density functional theory(DFT). We apply SMC-GPU to explore nanostructure evolution in two high-entropy alloys, FeCoNiAlTi and MoNbTaW, revealing diverse morphologies including nanoparticles, 3D-connected NPs, and disorder-stabilized phases. We quantify their size, composition,and morphology, and simulate an atom-probe tomography (APT) specimen for direct comparison with experiments. Our results highlight the potential of large-scale, data-driven MC simulations in exploring nanostructure evolution in complex materials, opening new avenues for computationally guided alloy design.
\end{abstract}


\end{frontmatter}


\section{Introduction}
\label{Introduction:1}
A central goal of computational materials science is to predict the physical properties of materials using only fundamental inputs such as atomic species and physical constants. A widely adopted strategy towards this goal is the direct solution of the quantum equations governing electron dynamics: the Schr\"odinger equation for non-relativistic cases and the Dirac equation for relativistic cases. However, the complexity of quantum many-body interactions necessitates approximations. For such a purpose, density functional theory (DFT) \cite{10.1063/1.4704546} has emerged as a particularly successful approach, which approximates intractable many-body effects through an exchange-correlation function within a single-body framework. This approximation significantly enhances computational efficiency, enabling the routine prediction of ground-state physical properties for systems with several hundred atoms. However, for systems requiring more than thousands of atoms to simulate, such as nanodefects, non-stoichiometric compounds, and complex solid-solution alloys, conventional DFT methods become prohibitively expensive due to their intrinsic $O(N^3)$ scaling behavior with system size. A further complication arises when considering finite-temperature effects \cite{MA201590}. Even for relatively small systems, rigorously accounting for temperature via ensemble averaging in statistical mechanics necessitates the evaluation of an enormous number of configurations, making DFT-based finite-temperature simulations prohibitively expensive. For example, a direct thermodynamic simulation of the CuZn alloy using a 250-atom supercell requires the calculation of 600,000 DFT energies \cite{PhysRevB.93.024203}, a task demanding some of the world’s largest supercomputers. 

One common strategy to accelerate atomistic simulations is to replace computationally expensive electronic structure calculations with efficient atomistic models. From the perspective of renormalization group theory \cite{RevModPhys.47.773, PhysRevB.99.134206, PhysRevB.29.4030}, the atomistic models can be interpreted as the effective model obtained by integrating out the electronic degrees of freedom, which substantially reduces the computational cost. Indeed, traditional empirical models, such as the embedded atom model (EAM) and cluster expansion methods, have been widely employed for large-scale thermodynamic simulations. However, traditional empirical models are limited in terms of functional form, which restricts their generalization capability and predictive accuracy. On the other hand, the advent of machine learning (ML) for atomistic systems \cite{GNNFF_NPJ_2021, NequIP_NC, MLPAlloys2024, 2025Nature} has introduced a new paradigm to the construction of effective atomistic models \cite{doi:10.1021/acs.jctc.2c01146, PhysRevLett.104.136403, SchnetPack, gasteiger_gemnet_2021, ying2021do}.
Instead of being predefined, the parameters in ML atomistic models are automatically determined by training from high-quality first-principles datasets, therefore rendering it possible for the model to automatically capture complex interatomic interactions \cite{ML_alloys, LIU2023101018, Zhang_2025, doi:10.1021/acs.jpcc.4c01704}. Well-trained ML models are typically orders of magnitude faster than DFT methods, while can still retain their high accuracy \cite{UniversalModel_NCS, doi:10.1126/science.abn3445, 2025Nature}. This renders it possible to carry out high-accuracy atomistic simulations using supercells containing millions of atoms \cite{10.5555/3433701.3433707, Nature_2021_Silicon, MPNNs_2023, 10.1145/3503221.3508425}, far exceeding the thousands-of-atoms limits of conventional DFT approaches. 

However, most machine learning potentials have been developed primarily to accelerate molecular dynamics (MD) simulations \cite{10.5555/3433701.3433707, Nature_2021_Silicon, NequIP_NC, MLPAlloys2024, doi:10.1021/acs.jctc.0c01343, doi:10.1021/acs.jcim.0c00451}, with relatively limited applications in Monte Carlo (MC) simulations \cite{Korman_npj, WANG2025120635, LIU2023101018, doi:10.1021/acs.jctc.4c00463, WANG2025120635}. This is notable given that MC represents one of the two cornerstone methods in atomistic simulations, alongside MD \cite{JIANG20231341}. A key challenge lies in the intrinsic sequential updating nature of widely used MC algorithms, such as the Metropolis algorithm. The MC trials are generally attempted site by site, which hinders large-scale parallelization. This is in stark contrast to MD, where all the atoms are updated simultaneously in a single step. The sequential updating nature of MC quickly becomes the bottleneck as the system size grows, which undermines the efficiency advantage of ML atomistic models when integrating the two. On the other hand, for many interesting phenomena \cite{e18080403, LIU2021110135, Junqi-NCS}, such as order-disorder transitions in chemically complex materials, MC simulations remain the only viable approach due to their sampling efficiency. Consequently, developing highly scalable MC algorithms that overcome the limitations of sequential updating is of paramount importance for
realizing the potential of ML atomistic models.

\begin{table}
    \centering
    \begin{tabular}{|>{\centering\arraybackslash}p{0.2\linewidth}|>{\centering\arraybackslash}p{0.15\linewidth}|>{\centering\arraybackslash}p{0.12\linewidth}|>{\centering\arraybackslash}p{0.12\linewidth}|>{\centering\arraybackslash}p{0.08\linewidth}|>
    {\centering\arraybackslash}p{0.15\linewidth}|} \hline 
         Model& Model Type &  Chip &  Distributed&  Year&  Journal \& Ref.\\ \hline 
         {Lennard-Jones}&  Pairwise&  CPU& \bf{Y}&  1986 & Thesis \cite{Johnson1986}\\ \hline 
         {Lennard-Jones}&  Pairwise&  CPU& \bf{Y}&  1996 & JCC \cite{1996-JCC}\\ \hline 
         Ising&  NN&  \bf{GPU}&  N&  2009 & JCP \cite{PREIS20094468} \\ \hline 
         Ising & NN & \bf{GPU} &\bf{Y}&2010 & CPC \cite{BLOCK20101549}\\ \hline
         {Cluster-expasnion} & \bf{Short-range} & CPU &\bf{Y}&2012 & PRB \cite{sadigh2012scalable}\\ \hline
         {Lennard-Jones}& Pairwise & \bf{GPU} & N & 2013 & JCP \cite{mick2013gpu}\\ \hline     
         Ising& NN& \bf{FPGA} & N & 2013 & JCP\cite{OrtegaZamorano2013FPGAIsing}\\ \hline
         Hard-disk& Hard core & \bf{GPU} & N & 2013 & JCP \cite{anderson2013massively}\\ \hline
         Hard-disk& Hard core &  \bf{GPU} &\bf{Y}& 2016 & CPC \cite{Anderson2016SMC}\\ \hline 
         {Coulomb} & {Long-range} & \bf{GPU} & N & 2017 & JCP\cite{Liang2017GpuMonteCarlo}\\ \hline  
         Ising & NN & \bf{TPU} &\bf{Y}& 2019 & 
         SC \cite{liu2019high}\\ \hline 
         Ising & NN & \bf{GPU-TC} & \bf{Y} & 2020 & CPC \cite{ROMERO2020107473} \\ \hline
         Magnetic & {Long-range} & \bf{GPU} &\bf{Y}& 2023 & JAP \cite{lepadatu2023accelerating}\\ \hline  
       {EPI, qSRO}&  \textbf{Short-range}&\textbf{GPU}& \textbf{Y}&{2025}& This work \\ \hline 
    \end{tabular}
    \caption{\color{black}Comparison of the method proposed in this work with previous checkerboard variants of parallel MC simulation, for atomistic simulation of materials. Algorithms generalizable to MLP, implemented on accelerators, or distributed over multiple chips, are highlighted with bold font. In the model type column, NN refers to nearest-neighbor.}
    \label{tab:MC}
\end{table}

{\color{black}
A widely adopted technique to address the parallelization challenge in Monte Carlo (MC) simulations is the checkerboard algorithm, originally developed for the Ising model with nearest-neighbor interactions. The combination of Ising model and checkerboard algorithm presents great parallel opportunities at core level, and has been successfully applied across various accelerators, including GPUs \cite{PREIS20094468,BLOCK20101549}, GPU tensor cores \cite{ROMERO2020107473}, TPUs \cite{liu2019high}, and ASICs \cite{OrtegaZamorano2013FPGAIsing}. Extending the checkerboard algorithm to non-Ising models poses greater challenges for achieving efficient parallelism. One of the earliest such efforts was by M. Johnson \cite{Johnson1986} for Lennard-Jones potential. Later, J. Anderson and collaborators proposed an extension for off-lattice hard-disk systems \cite{anderson2013massively, Anderson2016SMC}.
However, a key limitation of these efforts is their restriction to pairwise interactions, making them unsuitable for machine-learned potentials (MLPs), which often involve many-body interactions. For general short-range interactions, B. Sadigh and co-authors introduced the Scalable Parallel Monte Carlo (SPMC) algorithm \cite{sadigh2012scalable}, which supports higher-order interactions such as those in cluster expansion models. Their implementation was later integrated into the hybrid MC-MD framework in the LAMMPS package \cite{THOMPSON2022108171} and has been widely adopted for use with MLPs, including models like SNAP \cite{2020NPJ_Li} and MTP \cite{Yin2021}. Nevertheless, the original SPMC implementation was designed for CPU architectures, not GPUs. Moreover, it relies on a static domain decomposition scheme, which limits the parallel scalability. Finally, the practice of confining the local trial moves to non-interacting domains produces weak spatial correlations that can slow down the approach to equilibrium \cite{sadigh2012scalable}.
}

{\color{black}To overcome the above challenges, we introduce the Scalable Monte Carlo at eXtreme (SMC-X) method. Compared to previous methods summarized in Tab.~\ref{tab:MC}, SMC-X is unique in that it is designed for both two targets: (1) efficient implementation on accelerators, (2) support for arbitrary short-range interactions, such as those found in machine-learned potentials (MLPs). The key innovation in SMC-X is the introduction of the Local Interaction Zone (LIZ) within the link-cell (LC) strategy (a checkerboard variant), which enables efficient execution on accelerators. Although SMC-X shares conceptual similarities with the SPMC method \cite{sadigh2012scalable}, it avoids the key limitations of that approach mentioned above. In SMC-X, the LIZ+LC region is dynamically centered on the atoms updated in each mini-sweep, which brings two major advantages: (1) The parallelism scales linearly with the number of atoms, far exceeding the core-limited scalability of SPMC, and making it well-suited for high-throughput accelerators such as GPUs; (2) All sites are treated uniformly, maintaining equivalence with serial Monte Carlo updates (aside from update order, which does not affect detailed balance) and avoiding the spatial correlation issues inherent in SPMC. 
}

{\color{black}To demonstrate the capability of SMC-X in tackling challenging materials science problems previously intractable due to limitations in spatial and temporal scales, we employ SMC-X to study the nanoparticles in }high-entropy alloys (HEAs) \cite{ADEM:ADEM200300567, CANTOR2004213}. HEA is a class of chemically complex materials that have received significant attention due to their exceptional mechanical properties. These properties include overcoming the traditional strength-ductility trade-off \cite{NatureRaabe, Yang933, ORNL_HEAs_2021}, attributed to phenomena such as chemical short-range order \cite{SRO_NCS_2023, Junqi-NCS, APT_CoCrNi_NatureMat_2024}, nanoprecipitates \cite{Yang933}, and nanophases \cite{NanoParticleHEA2022}. Understanding these features necessitates large-scale MC simulations. Beyond mechanical properties, the size and morphology of nanostructures in HEAs also offer promising opportunities for catalysis, sparking widespread interest in recent years \cite{HAN20231717, doi:10.1126/sciadv.adn2877, doi:10.1126/science.abn3103}. 
{\color{black} As a general algorithm, SMC-X can be applied to different accelerators. In this work, we tailored the parallelization strategies in SMC-X for efficient execution on GPUs, and refer to this implementation as SMC-GPU. Specifically, we apply SMC-GPU} to investigate the nanostructure evolution in the $\rm{Fe_{29}Co_{29}Ni_{28}Al_7Ti_7}$ and MoNbTaW HEAs, employing two simple machine learning energy models using the local short-range order parameters \cite{ZHANG2020108247, LIU2021110135, doi:10.1021/acs.jctc.4c00340} as input features.
We demonstrate the excellent efficiency and scalability of the SMC-X method, which enables the simulation of atomistic systems exceeding one billion atoms using a single GPU (graphic processing unit). Our results reveal a rich diversity of interesting nanoscale phenomena, including nanoparticle (NP), 3D-connected NPs, and disorder protected nanophases. We quantitatively analyze the size, composition, and morphology of the nanostructures, which align well with available experimental results obtained with atom-probe-tomography (APT) and electron microscopy. Finally, our results reveal that the intricate nanoscale interplay of order and disorder in high-entropy alloys (HEAs) stems from the combined effects of chemical complexity and temperature, offering valuable guidance for alloy design.

\section{Results}
\subsection{Performance of the SMC-X algorithm}
\label{Results:SMC}
\begin{figure} [ht!]
    \centering
    \includegraphics[width=1.0 \linewidth]{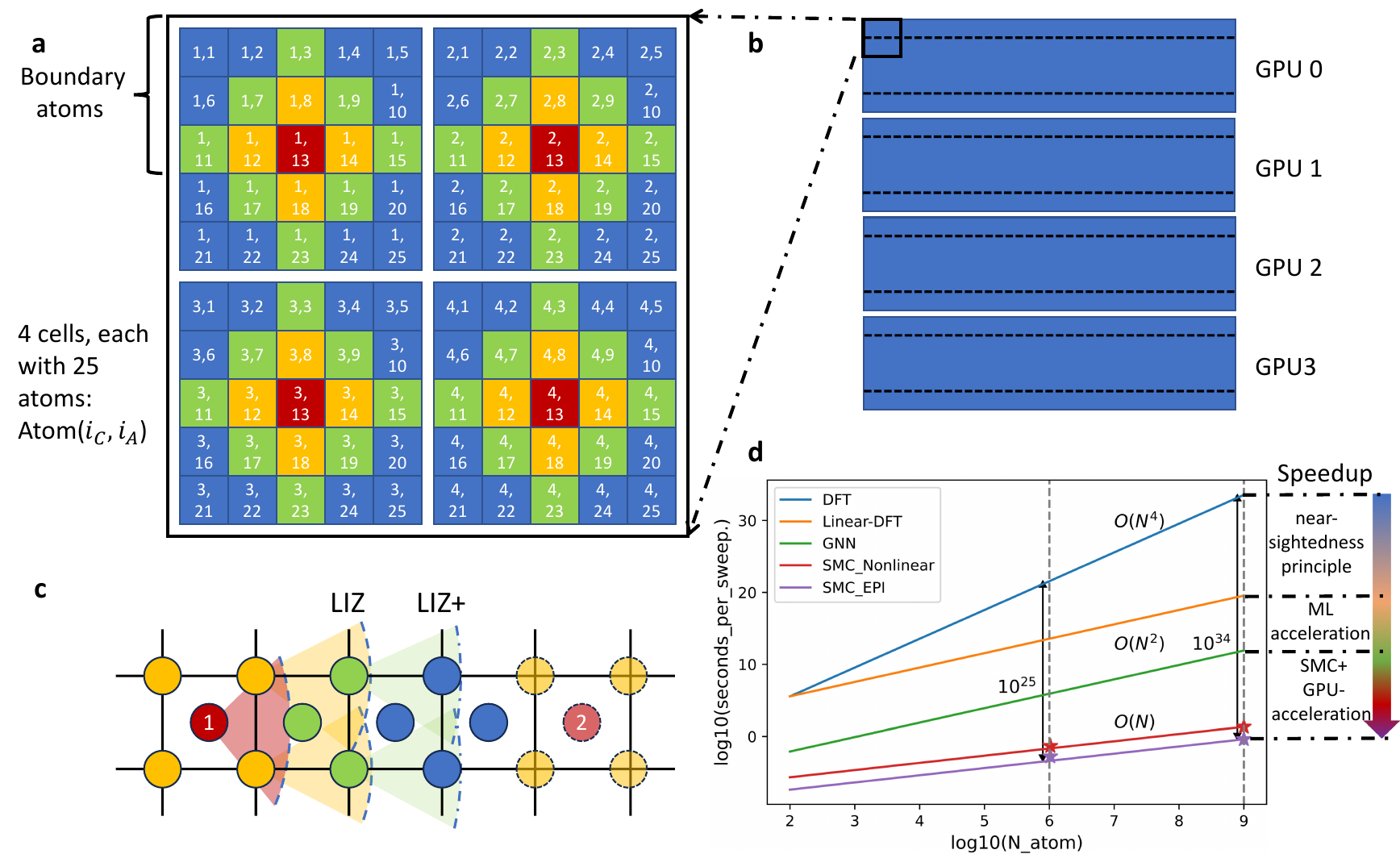}
    \caption{Schematics of the scalable Monte Carlo (SMC-X) method. (a) Illustrate the SMC-X method with a 2D square lattice, in which each site is denoted by two indices $(i_C, i_A)$, where $i_C$ represents different link-cells, and $i_A$ represents the different atoms within the $n_A$ atom cell. (b) The whole chemical configuration is distributed on multiple GPUs, and the atoms near the boundary need to be communicated between GPUs $n_A$ times in every MC sweep. (c) A 2D view of the fcc or bcc lattice, site 1 (red) can swap with each of its nearest neighboring sites (yellow), and the green sites represent the ones that the local energies can be affected. The yellow and green sites, together with the centering atom form the LIZ. The chemical environment with LIZ+ are needed to calculate the energy change due to a swap trial. (d) A log-log plot of computation time vs the system size to illustrate the speed-up ratio of the SMC-X method as compared to DFT (MuST-KKR), linear-DFT (LSMS), and GNN (Allegro). The measured values for SMC-X are signified as stars and ideal scaling is assumed for all lines.}
    \label{fig:link-cell}
\end{figure}

The SMC-X algorithm can be seen as a generalization of the checkerboard algorithm for the 2D Ising model, as explained in Method~\ref{Method:SMC-GPU} and Supplementary~\ref{S:1}. The key insight lies in rendering sequential Monte Carlo (MC) trials independent by partitioning the system into sufficiently large link-cells, thereby isolating the influence of individual MC trial moves. To simplify the explanation, a schematic of the link-cell for the 2D square lattice is shown in Fig.~\ref{fig:link-cell} (a). Each site is labeled by two indices $(i_C, i_A)$, where $i_C$ represents different link-cells and $i_A$ represents the different atoms within the link-cell. For the sake of discussion, let us assume that the atoms have only nearest-neighbor interactions. In such a case, the MC updates of the atoms with the same $i_A$ index but different $i_C$ indexes are independent of each other. This obviously presents a parallelism opportunity. For instance, consider atom $(1,13)$. During MC update, atom $(1,13)$ can swap with any of its nearest neighbors $(1,12)$, $(1,14)$, $(1,8)$ and $(1,18)$, which are colored as yellow in Fig.~\ref{fig:link-cell}(a). Due to the short interaction range (nearest-neighbor), these moves would have no impact on calculating the energy changes for the swap moves of atom $(i_C,13)$ (red sites in Fig.~\ref{fig:link-cell}(a)), and the degree of parallelism is the number of cells $n_C$. We denote the yellow and green sites surrounding each red site in Fig.~\ref{fig:link-cell} as a local-interaction zone (LIZ), which is a name inspired by the locally self-consistent multiple scattering (LSMS) method \cite{PhysRevLett.75.2867, osti_1420087}. In other words, for any two sites, as long as the MC trial of one site does not affect the energies of the LIZ of the other site, then these two sites are independent. It is easy to see that the above discussion based on the 2D square lattice can be extended to the case of 3D crystal. Furthermore, in the SMC-X method, a domain decomposition scheme is introduced to distribute the lattice among multiple GPUs, which further enhances the achievable system size, as illustrated in Fig.~\ref{fig:link-cell} (b).
Generally, to determine the energy change resulting from the MC swap trial at site $i$, the local energies of each site within the local interaction zone of site $i$ must be evaluated. Additionally, chemical environment information extending beyond the LIZ is required for the energy change calculation, as illustrated by the LIZ+ region in Fig.~\ref{fig:link-cell}
(c). Note that for a pair-interaction model, the above discussion can be simplified since the total energy changes can be calculated from the local energies of the two sites involved in the swap trial. As a result, the speed of the effective pair-interaction (EPI) model is faster than the generalized nonlinear model by approximately a factor of 50, as shown in Fig.~\ref{fig:link-cell} (d). 

To illustrate the strength of our method, we compare the theoretical speed of the SMC-X method with other schemes to integrate MC with high-accuracy energy prediction methods, such as DFT, linear-DFT, and GNN (graph neural networks), as shown in Fig.~\ref{fig:link-cell} (d). Note that the lines are drawn by extrapolating from a data point using ideal scaling behavior. The actual computational speeds of the various methods are inherently influenced by many factors, such as hardware, software implementations, material systems, and computational parameters. Therefore, the discussion presented here is intended solely to provide a theoretical estimation of their relative performance in terms of orders of magnitude. The stars in the figure represent the measured values of the SMC-X method, and the lines show the ideal scaling behaviors, which are explained in Method~\ref{Method:scaling}. From Fig.~\ref{fig:link-cell} (d), we see that for a one million atom system, we have the speedup $S = 10^{25}$, as highlighted in  Fig.~\ref{fig:link-cell} (d). We can also see that for a nonlinear model, such as the {\color{black} qSRO} model introduced in the following section, the speedup will be reduced by a factor of 50, due to the aforementioned necessity of evaluating the local energies of each site within the LIZ.
If we increase the system size to one billion atoms, then the speedup of SMC-X with respect to DFT will be further increased by a factor of $10^9$ to reach $10^{34}$, as highlighted in Fig.~\ref{fig:link-cell} (d). This analysis clearly demonstrates the exceptional performance of the SMC-X method for large-scale MC simulations.

In addition to theoretical estimations, we compare our work with selected previous studies, as summarized in Tab.~\ref{tab:methods}. It can be seen that our work represents the largest system size achieved in atomistic simulations at ab initio accuracy. Such capability enables us to directly study large nanostructures comprising of millions of atoms, in contrast to previous works that are limited to short-range order (SRO), long-range order (LRO), and small nanoprecipitates.
To contextualize the simulation scale of our work, we extend our search beyond HEAs to identify the largest atomistic simulation system reported in the literature. A recent study achieved this with a system of 29 billion atoms, which employs 35 million CPU cores on one of the world's largest supercomputers (comprising 90,000 computing nodes, approximately 84\% of the entire system). In contrast, constrained by computational resources, our work utilized only two NVIDIA H800 GPUs, yet already achieved a simulation scale of one billion atoms. 
Note that this already surpasses the 100-million-atom system size of the 2020 Gordon-Bell Prize winner \cite{10.5555/3433701.3433707}. Furthermore, the comparison presented above demonstrates that, through careful design of the Monte Carlo algorithm to fully exploit its inherent parallelization potential, MC simulations can rival molecular dynamics (MD) methods in terms of scalability, which are traditionally viewed as advantageous over MC methods in simulation scale. Finally, we highlight that the excellent computational efficiency of the SMC-X method not only presents a success in pushing the simulation length scale forward, but also provides a particularly important tool for understanding the vital role of nanostructures \cite{NanophaseHEA, ORNL_HEAs_2021} in the exceptional mechanical properties in HEAs \cite{doi:10.1126/science.abn3103, Fueaat8712, Yang933}, such as overcoming the trade-off in strength and ductility \cite{NanophaseHEA, Yang933}.

\begin{table}
    \centering
    \begin{tabular}{|>{\centering\arraybackslash}p{0.17\linewidth}|>{\centering\arraybackslash}p{0.15\linewidth}|>{\centering\arraybackslash}p{0.15\linewidth}|>{\centering\arraybackslash}p{0.10\linewidth}|>{\centering\arraybackslash}p{0.3\linewidth}|>{\centering\arraybackslash}p{0.08\linewidth}|} \hline 
         Materials&  Simulation Method&  Model&  System Size (Atoms)&  Subject&  Ref.\\ \hline 
         MoNbTaW, MoNbTaWV, MoNbTaWTi&  Canonical MC&  EPI (6th-NN)&  1,000&  SRO, order-disorder transition& \cite{LIU2021110135}\\ \hline 
         AlCoCrFeNi&Canonical MC&EPI (NN)& 6,912&LRO and SRO, ordered multi-phase HEA&\cite{Santodonato2018}\\ \hline 
         NiCoFeAlTiB&  Semi-grand canonical MC& EPI (4th-NN)&   32,000&   Phase diagram, SRO and LRO &\cite{WANG2025120635}\\ \hline 
         NbMoTaW &  Hybrid MC/MD&  SNAP&  36,000&  Grain boundaries (GBs) and SRO&\cite{2020NPJ_Li}\\ \hline 
         MoNbTaW& Hybrid MC/MD& MTP& 573,672& Dislocation motion under SRO& \cite{Yin2021}\\ \hline  

         MoNbTaW & Hybrid MC/MD & SNAP & 580,000 & Dislocation motion under SRO and nanoscale B2 precipitates & \cite{YAO2024120457}\\ \hline
       \textbf{FeCoNiAlTi, MoNbTaW}&  \textbf{Canonical MC}&\textbf{EPI (2nd-NN) + nonlinear ML model}& \textbf{$1\times 10^9$}&\textbf{ Nanostructures (size, composition, and morphology)}& \textbf{This work} \\ \hline
       Water/Copper Systems& MD& DeepMD& $29\times 10^9$ & Performance benchmark on an exascale supercomputer with 35-M CPU cores&\cite{10880101}\\ \hline
    \end{tabular}
    \caption{Comparison of our work with previous studies employing atomistic simulations for high entropy alloys. For reference, we also include the largest atomistic system (29-B) simulated using machine learning (ML) models, as identified in the literature. Notably, Ref.~\cite{10880101} utilized 35 million CPU cores in one of the world's fastest supercomputers, whereas this work achieved a billion atoms using only two NVIDIA H800 GPUs.}
    \label{tab:methods}
\end{table}

\subsection{Energy model}
\begin{figure} [ht!]
    \centering
    \includegraphics[width=0.95 \linewidth]{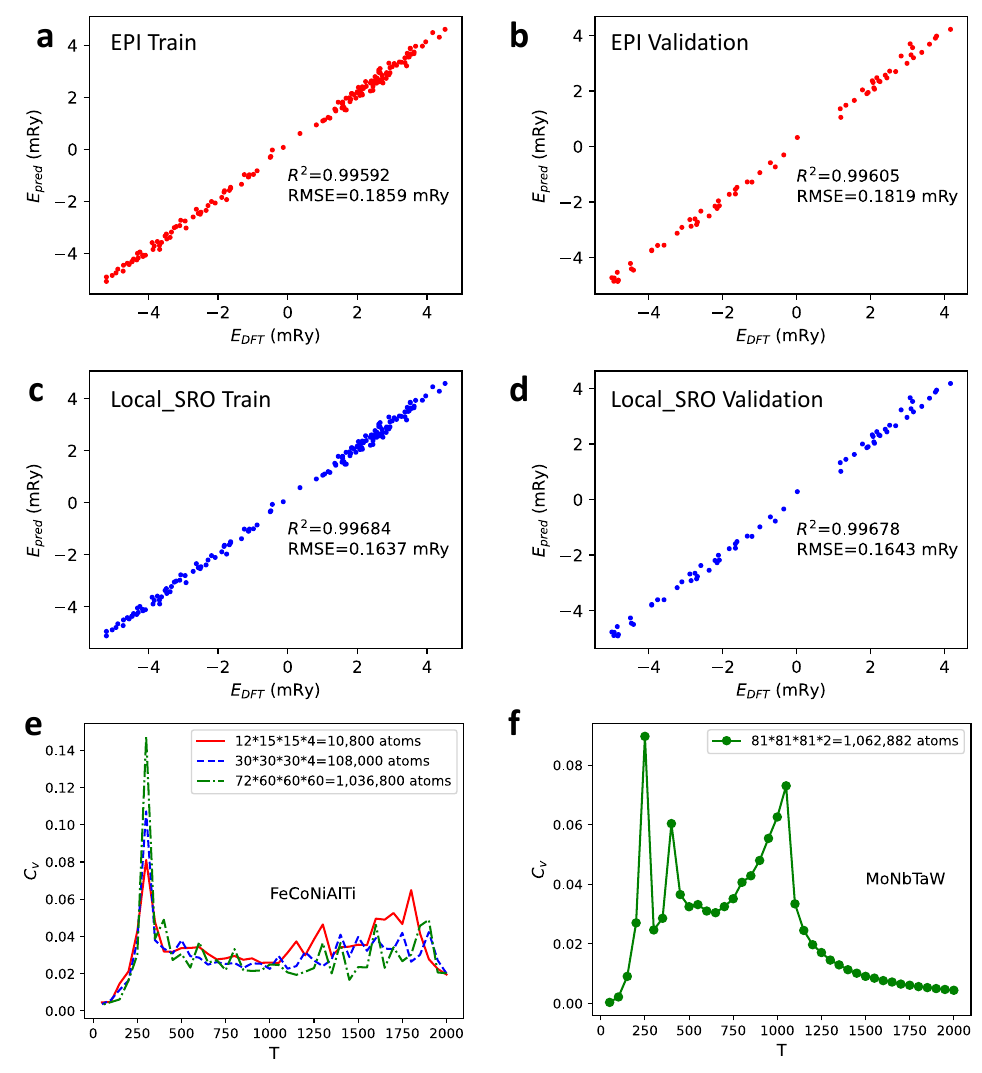}
    \caption{The accuracy of the energy model {\color{black}(RMSE per atom)} and the calculated specific heats for FeCoNiAlTi and MoNbTaW. (a) The training error of the EPI model for FeCoNiAlTi. (b) The validation error of the EPI model for FeCoNiAlTi (0.7/0.3 train-valid splitting). (c) The training error of the loca\_SRO model for FeCoNiAlTi. (d) The validation error of the qSRO model for FeCoNiAlTi.  (e) The specific heats $C_V$ of FeCoNiAlTi were calculated with different supercell sizes.  (f) The specific heats of MoNbTaW were calculated with a one-million-atom supercell.}
    \label{fig:CV}
\end{figure}

In this work, we focus on two types of energy models: effective-pair-interaction (EPI) model \cite{liu2019machine, ZHANG2020108247}, and a nonlinear model that uses the local SRO parameter as the input feature. The EPI model is a generalized Ising model constructed via machine learning from the DFT data. By automatically selecting the interaction range via Bayesian information criterion (BIC) \cite{ZHANG2020108247}, the EPI model has demonstrated that it can predict the DFT-calculated configuration energies of a series of HEAs with very high accuracy \cite{LIU2021110135}. The local-SRO model, proposed in this work, generalizes the EPI model by adding a quadratic term. As will be shown in the following discussion, this simple nonlinear term has a profound impact on the computing pattern, and serves as a prototypical representation for machine learning models. A detailed description of the two energy models are present in Method~\ref{Method:model}.

Using a train-validation splitting of 70\% and 30\% for the DFT dataset described in Method.~\ref{Method:DFT_data}, we evaluated the accuracy of the second-nearest-neighbor (2nd-NN) EPI model and the loca\_SRO model and the results are shown in Fig.~\ref{fig:CV} (a-d). Both models are trained with the Adam optimizer and back-propagation algorithm. It can be seen that both of the two models demonstrate very high accuracy. For the EPI model, the validation root-mean-square-error (RMSE) is {\color{black}0.1819 mRy/atom}, or approximately {\color{black}2.5 meV/atom}, which is well
within quantum chemical accuracy (approximately 3.16 mRy/atom), as well as smaller than the typical errors of $5-12 $ meV/atom in the DFT methods \cite{OQMD}. The qSRO model demonstrates slightly higher accuracy than the EPI model, with a validation RMSE of 0.1643 mRy/atom, which can be attributed to its inclusion of high-order interactions. The high accuracy of the models is also demonstrated in the $R^2$ scores, which are all higher than 0.995. The errors from the train and validation datasets are also very close to each other, indicating the absence of overfitting in the models. For comparison, we notice that this accuracy is higher than that in Ref.~\cite{WANG2025120635}, which is reported as 14.7 meV/atom by applying a similar pair-interaction model in the FeCoNiAlTiB system. We think one reason for the enhanced accuracy is our approach to obtaining the DFT dataset, which combines both random configurations, as well as configurations from Monte Carlo simulations, as proposed in Ref.~\cite{LIU2021110135} and described in Method.~\ref{Method:DFT_data}. 

Using the trained models, we calculated the specific heats $C_V$ in the FeCoNiAlTi HEA in temperatures ranging from 50 K to 2000 K. The calculation details are given in Method.~\ref{Method:MC}, and the results are shown in Fig.~\ref{fig:CV} (e). It can be seen that there is a sharp order-disorder transition at a low temperature of about 300 K. The sharp peak evolves to a singularity with the increase of the system size, which signifies the occurrence of a first-order transition. We also calculated the $C_V$ curve for the MoNbTaW refractory HEA. The results in Fig.~\ref{fig:CV} (f) agree well with the results in Ref.~\cite{LIU2021110135}, which are calculated using a small system of 1000 atoms. Note that the order-disorder transition at lower temperatures (between 250 K and 500 K) in the $C_V$ curve generally cannot be directly compared with experiments due to kinetic barrier effects. On the other hand, the results at elevated temperatures are more suitable for comparison if experimental data are available, as demonstrated in Ref.~\cite{Junqi-NCS}. A comparison of Fig.~\ref{fig:CV} (e) and (f) in the temperature range between 500 K and 2000K shows an interesting difference between the two HEAs: 
The $C_V$ curve in $\rm{Fe_{29}Co_{29}Ni_{28}Al_{7}Ti_{7}}$ is generally flat but demonstrates clear fluctuation as the temperature changes, indicating the existence of complex second phases or short-range order. By comparison, the $C_V$ curve of MoNbTaW in the same temperature region is smooth, but with an obvious order-disorder transition around 1000 K. Note that the exact location of the order-disorder transition temperature depends on the energy model and different values, as having been extensively reported in various theoretical works \cite{Huhn2013, Korman_npj, LIU2021110135, ZHANG2022105780}. 

\subsection{Nanostructures in $\rm{Fe_{29}Co_{29}Ni_{28}Al_7Ti_7}$}
\begin{figure} [ht!]
    \centering
    \includegraphics[width=1.0 \linewidth]{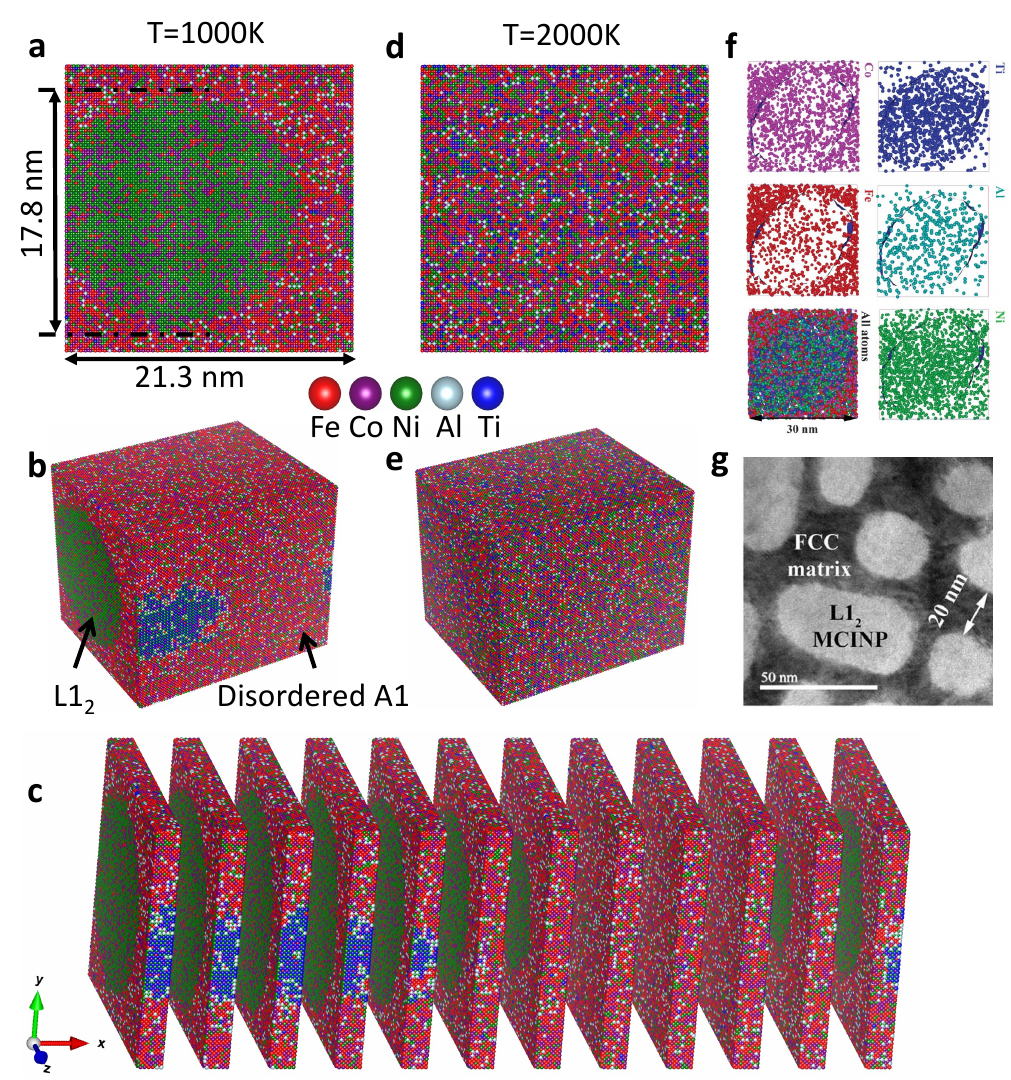}
    \caption{Simulation results for FeCoNiAlTi with one-million atoms. (a) A snapshot of the (001) face at 1000 K. (b) a perspective 3D snapshot of the configuration at 1000 K, from the signified direction (c) Slices of the atomic system to show the nanoparticle, using a step of 10 fcc layers. (d) A snapshot view of the (001) face at 2000 K (e) a 3D view of the configuration at 2000 K. (f) High-resolution atom maps showing the atomistic distribution within the $L1_2$ nanoparticles of the Al7Ti7 alloy. Reproduced from Fig.~2 in Ref.~\cite{Yang933}. (g) TEM image of the Al7Ti7 alloy showing the nanostructured morphology.  Reproduced from Fig.~1 in Ref.~\cite{Yang933} with reprint permission from Science. }
    \label{fig:FeCoNiAlTi_1M}
\end{figure}
\begin{figure} [ht!]
    \centering
    \includegraphics[width=1.0 \linewidth]{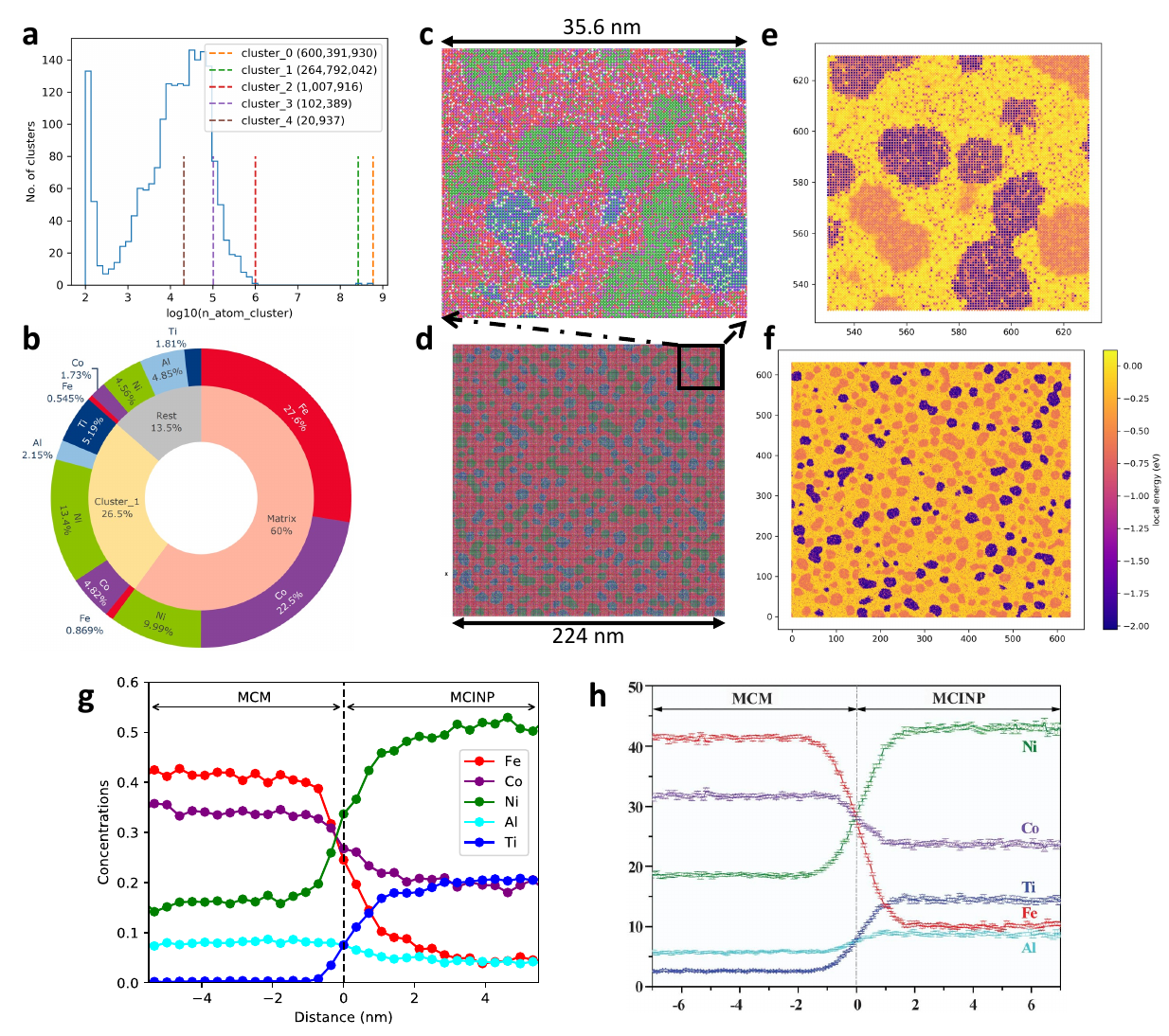}
    \caption{ Comparing the simulated configurations of the 1 billion atom FeCoNiAlTi with the experimental results. (a) A histogram for all clusters of sizes larger than 100 atoms. Five clusters of different sizes are selected and shown as vertical dashed lines. (b) The chemical concentrations of the matrix (cluster\_0) and cluster\_1, as well as the rest clusters. (c-f) Snapshots of the [001] face of NiCoFeAlTi at T=1000K, with (c) as the enlarged upper-right corner of (d), and (e) as the enlarged upper-right corner of (f). (c-d) show the different elements, and (e-f) show the atomic local energies. (g) Compositions of the NP (MCINP) and matrix (MCM) from the 1M atom simulation via sampling along the x direction using a radius of 6.4 nm. (h) Compositions of the NP (MCINP) and matrix (MCM) from experiment, as reproduced from Ref.~\cite{Yang933} with permission from Science.} 
    \label{fig:FeCoNiAlTi_1B}
\end{figure}

In this section, we report the nanostructure evolution of the $\mathrm{Fe_{29}Co_{29}Ni_{28}Al_7Ti_7}$ fcc HEA. This alloy has demonstrated exceptional combined strength and ductility in experiments, and such an attractive property has been attributed to the formation of nanoparticles in the FeCoNi matrix \cite{Yang933}, which hinder the motion of dislocations. In the experiment, the alloy sample is homogenized at 1150 $^\circ C$ (1423 K) for 2 hours and aged at 780 $^\circ C$ (1053 K) for 4 hours \cite{Yang933}. Using the local SRO model, we first simulated the chemical phase changes of the $\mathrm{Fe_{29}Co_{29}Ni_{28}Al_7Ti_7}$ fcc alloy using a supercell of approximately one million atoms. The lattice dimension is chosen as $72\times60\times60$, therefore, the total number of atoms is $72\times60\times60\times4=1,036,800$. We initialize the configurations randomly at 2000 K, then decrease the simulation temperature by 50 K each time until 1000 K, which is approximately the temperature at which the samples are aged.
At each temperature, we skip the $4\times10^5-1.5\times 10^6$ MC sweeps before recording the following 20,000 sweeps to calculate the specific heat $C_V$ in Fig.~\ref{fig:CV} (e). The configurations at 2000 K and 1000 K are demonstrated in Fig.~\ref{fig:FeCoNiAlTi_1M} (a-e). From Fig.~\ref{fig:FeCoNiAlTi_1M} (d-e) it can be seen that the system is chemically disordered at 2000 K. As the temperature decreases to 1000 K, a nanoparticle (NP) forms in the system. The distribution of elements in the NP, as shown in Fig.~\ref{fig:FeCoNiAlTi_1M} (a-b), reveals the formation of {\color{black}$\text{L1}_2$} structure. The two different layers of the $\text{L1}_2$ nanoparticle can be seen from Fig.~\ref{fig:FeCoNiAlTi_1M} (b-c). This agrees well with the experimental results in Ref.~\cite{Yang933}, as reproduced in Fig.~\ref{fig:FeCoNiAlTi_1M} (f-g), in which the nanoparticles are refered to as multi-component intermetallic nanoparticles (MCINP). Similar results are also reported in Ref.~\cite{FU2019372} for the $\rm{Fe_{25}Co_{25}Ni_{25}Al_{15}Ti_{10}}$ HEA, where the fcc phase consisted of a $\gamma$ Fe-(Co,Ni)-based solid-solution matrix (A1), and coherent primary $\gamma'$ (Ni,Co)3-(Ti,Al)-based intermetallic $\text{L1}_2$ precipitates. Other than chemical structure, size is another important feature of the nanoparticles. The diameter of the cluster shown in Fig.~\ref{fig:FeCoNiAlTi_1M} (a) is about 17.8 nm, which is in agreement with the experimental observation shown in Fig.~\ref{fig:FeCoNiAlTi_1M} (f-g). Moreover, we extract the compositions from the simulation data and show it in Fig.~\ref{fig:FeCoNiAlTi_1B} (g), along with the experiment results in Ref.~\cite{Yang933}, as shown in Fig.~\ref{fig:FeCoNiAlTi_1B} (h). It can be seen that the compositions from MC simulation generally agrees well with experiment, although there are also some differences, such as slightly higher concentration of Ni in MCINP.

\begin{figure} [ht!]
    \centering
    \includegraphics[width=1.0 \linewidth]{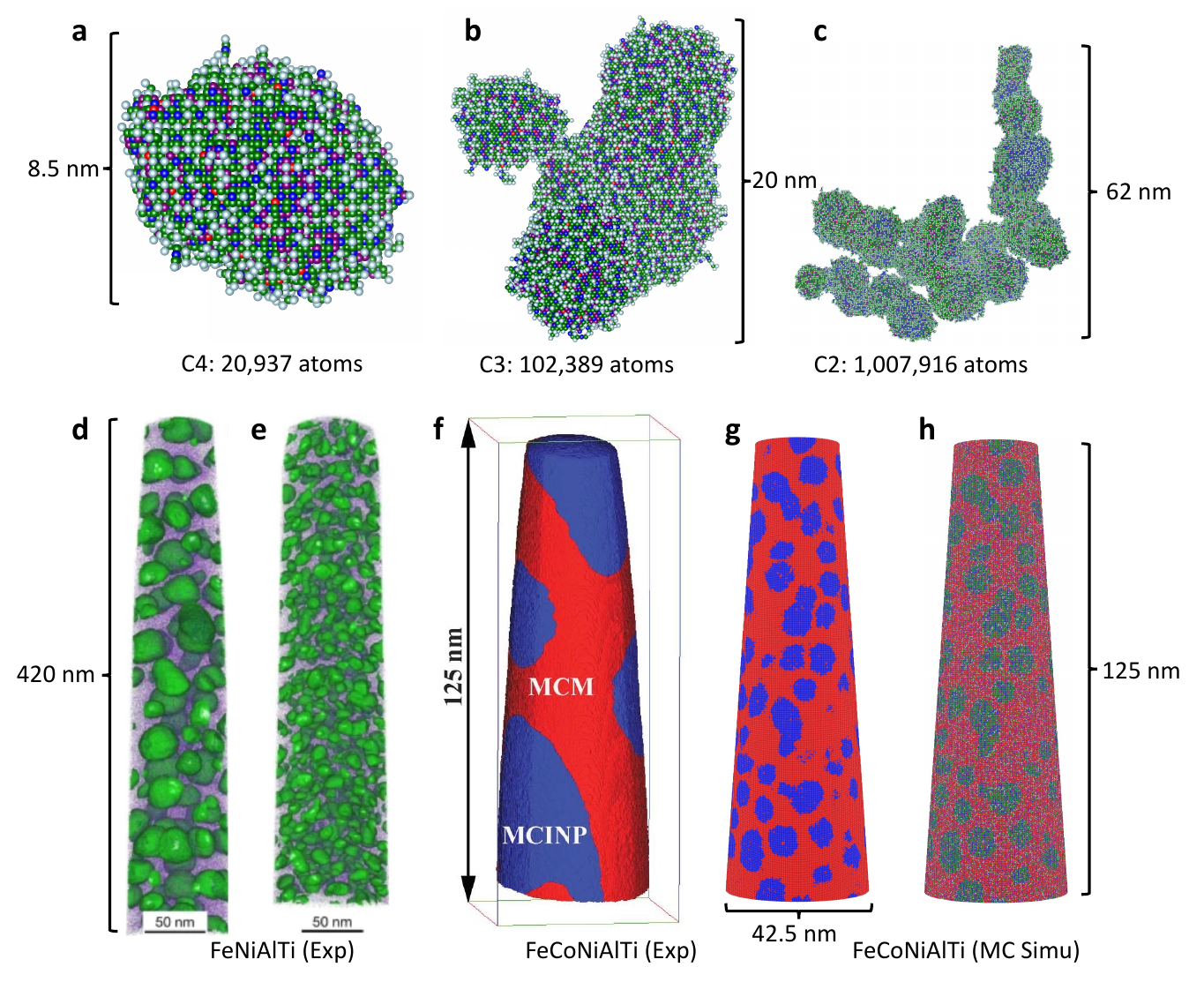}
    \caption{ Comparison between the NPs from MC simulation and APT experiment. (a-c) 3D snapshot of cluster\_4, 3, and 2, respectively. (d) APT results from fcc grains of FNAT sample (47-hours aging at 973 K) showing the distribution and composition of $\text{L1}_2$ precipitates in fcc matrix. (e) Same as (d), but for a sample obtained via 4-hours aging. Both (d) and (e) are reproduced from Ref.~\cite{ORNL_HEAs_2021}. We will seek permission once this manuscript being accepted. (f) The APT results of the $\mathrm{Fe_{29}Co_{29}Ni_{28}Al_7Ti_7}$ HEA, as reproduced from \cite{Yang933}. (g) The simulation result obtained using SMC-X. The blue region show the NP identified, and the red region show the matrix phase. (h) The simulation result obtained with SMC-X. The chemical species are explicitly shown. }
    \label{fig:FeCoNiAlTi_APT}
\end{figure}

From Fig.~\ref{fig:FeCoNiAlTi_1M} (b,c), we see that the one-million-atom supercell can only accommodate one NP in the FeCoNiAlTi HEA, and larger systems are necessary for a direct comparison of experimental data that contains more than one NP, such as the TEM image in Fig.~\ref{fig:FeCoNiAlTi_1M} (g). To investigate the microstructures of the FeCoNiAlTi HEA at a larger length scale, we employ a $630\times630 \times 630 $ fcc supercell, in which the total number of atoms is 1,000,188,000. Due to the huge supercell size, we use the faster EPI model rather than the local SRO model, and the simulation details are specified in Method.~\ref{Method:MC}. The simulation results for this one billion-atom system are presented in Fig.~\ref{fig:FeCoNiAlTi_1B} and Fig.~\ref{fig:FeCoNiAlTi_APT}. To study the size, composition, and morphologies of the nanoparticles, we employ the union-find algorithm to identify the nanoparticles, as detailed in Method.~\ref{Method:MC}. We refer to all sets of atoms identified by the union-find algorithm as clusters, which can be of sizes ranging from a single isolated atom to the whole matrix phase. We make a histogram of all clusters of sizes larger than 100 atoms in Fig.~\ref{fig:FeCoNiAlTi_1B} (a). The largest one is cluster\_0, which is the matrix phase. Cluster\_0 has a total of 0.6 billion atoms, and contains only Fe, Co, and Ni, as demonstrated in Fig.~\ref{fig:FeCoNiAlTi_1B} (b). Note that the absence of Al and Ti atoms in cluster\_0 is a result of the union-find algorithm, which automatically identifies isolated Al and Ti atoms in cluster\_0 as individual clusters of size one. From Fig.~\ref{fig:FeCoNiAlTi_1B} (c), we see that the FeCoNi matrix phase is a disordered A1 structure. The cause of the disorder can be seen from Fig.~\ref{fig:FeCoNiAlTi_1B} (e) and (f), which show the local energies of each atom with the color bar. It can be seen that in the matrix phase, the local energies of different atoms are generally close to each other, which means that the Fe, Co, and Ni atoms have no significant site preferences. On the other hand, the nanoparticles, which forms $\text{L1}_2$ structure as shown in Fig.~\ref{fig:FeCoNiAlTi_1B} (c), contains sites with much lower atomic local energies, as shown in Fig.~\ref{fig:FeCoNiAlTi_1B} (e). In fact, the local energy difference is exactly the criteria we make use of to distinguish the matrix and nanoparticles, as described in Method~\ref{Method:MC}. The FeCoNi matrix phase is the entropy-stablized disordered phase, while the $\text{L1}_2$ phase is the enthalpy favorable ordered phase. In other words, the occurrence of both disordered A1 structure and ordered $\text{L1}_2$ structure is a result of the competition between the entropy and enthalpy contributions in the free energy.

To study the size, compositions, and morphologies of the nanoparticle in more detail, we choose four representative nanoparticles, as illustrated in Fig.~\ref{fig:FeCoNiAlTi_1B} (a) as cluster\_1, 2, 3, and 4, which have 264,792,042, 1,007,916, 102,389, and 20,937 atoms. Their chemical concentrations are also listed in Tab.~\ref{tab:clusters}, which aligns well with the experimental results \cite{WANG2025120635}. From Fig.~\ref{fig:FeCoNiAlTi_1B} (a), we note that the peak of the histogram is at about $10^5$ atoms, which corresponds to a cluster of 29 atoms, or 10.3 nm in diameter. For cluster\_1, a similar calculation gives a value of 22.0 nm in diameter, if we assume that the particles are spherical. 
However, inspecting the 3D morphology of the clusters reveals that they are actually not simply spheres or ellipsoids, as shown in Fig.~\ref{fig:FeCoNiAlTi_APT} (a-c): While ellipsoid is a good approximation for the smaller cluster\_4, the larger cluster\_3 and cluster\_2 are both comprising of multiple nanoparticles, which indicates that the smaller nanoparticles can merge together to form larger ones, as shown by the dendritic structures at the surface of the nanoparticles. In fact, from Fig.~\ref{fig:FeCoNiAlTi_1B} (a,b) we see that the largest $\text{L1}_2$ cluster: cluster\_1, contains a total of more than 264 million atoms, which accounts for 66\% of the atoms in the NP phases. In other words, most of the nanoparticles shown in Fig.~\ref{fig:FeCoNiAlTi_1B} (c-f) are actually connected at 3-dimension and are parts of cluster\_1. Compared to simple NP, we propose that this 3D-connected-NPs (3DCNP) can further hinder the motion of dislocations, which can serve as a strengthening mechanism in these types of HEAs. Verifying the existence of the 3DCNP demands high-precision experimental techniques such as atom probe tomography (APT). The APT image of $\rm{Fe_{29}Co_{29}Ni_{28}Al_7Ti_7}$ is available in Ref.~\cite{Yang933}, as reproduced in Fig.~\ref{fig:FeCoNiAlTi_APT} (f). However, the image only shows the surface of the sample instead of the 3D chemical structure. Although the theoretical observation of the 3DCNP still requires experimental validation, we contend that its existence is plausible, given the high density of NPs observed in experiment \cite{Yang933}. Moreover, the APT result of the FNAT alloy \cite{ORNL_HEAs_2021}, a medium entropy alloy made up of the Fe, Ni, Al, and Ti elements, seems to support the existence of $\text{L1}_2$ 3DCNP, as shown in Fig.~\ref{fig:FeCoNiAlTi_APT} (d,e). For a direct comparison with the experimental results, we also show the simulation results of the APT sample needle in Fig.~\ref{fig:FeCoNiAlTi_APT} (g, h). It is easy to see that the general shapes of the NPs from simulation resemble the experimental results of FNAT, despite their different chemical compositions. The sizes of the NPs from simulation is generally smaller than experimental results, which could be due to the relatively limited $10^5$ MC simulation steps. Again, the importance of SMC-X method is highlighted by the capability to directly simulate an APT needle comprising of 10 million atoms. Other than the mechanical properties, the NPs can also have important application in catalysis \cite{3DCNP-ACS}, which would be an interesting topic for future research \cite{doi:10.1126/sciadv.adn2877, HEA_Nano_ACS_Nano}.

\begin{table}
    \centering
    \begin{tabular}{|c|c|c|c|c|c|} \hline  
         &  Ni&  Co&  Fe&  Al& Ti\\ \hline  
         L1$^{EXP}_2$&  43.23&  23.69&  10.06&  8.61& 14.41\\ \hline  
         Matrix$^{EXP}$&  18.69&   31.84&  41.13&  5.69& 2.66\\ \hline 
 L1$^{Ref}_2$& 49.86& 24.05& 1.60& 11.15&13.48\\\hline
 Matrix$^{Ref }$ & 2.34& 33.51& 66.53& 1.19&0.44\\\hline 
         Cluster$^{MC}_1$&  50.8&  18.2&  3.28&  8.11& 19.61\\ \hline  
         Cluster$^{MC}_2$&  50.77&  18.26&  3.32&  8.01& 19.63\\ \hline  
         Cluster$^{MC}_3$&  50.72&  18.25&  3.28&  8.15& 19.6\\ \hline  
         Cluster$^{MC}_4$&  50.51&  17.92&  3.38&  8.55& 19.64\\ \hline 
          Matrix$^{MC}$& 16.64& 37.41& 45.96& 0.0&0.0
\\ \hline 
    \end{tabular}
    \caption{
    Comparison of the experimental values at 1055 K and theoretical values at 1100 K from Ref.~\cite{WANG2025120635} with our simulation results at 1000 K for the compositions of the matrix and four representative $\rm{\text{L1}_2}$ clusters, as listed in Fig.~\ref{fig:FeCoNiAlTi_1B}. Note that $\rm{Matrix}^{MC}$ is Cluster\_0. Due to the adoption of the union-find algorithm, the isolated atoms are automatically removed from the identified clusters, which is the cause of no Al and Ti atoms in the extracted matrix phase.}
    \label{tab:clusters}
\end{table}

\subsection{Nanostructures in MoNbTaW}
\begin{figure} [ht!]
    \centering
    \includegraphics[width=0.8 \linewidth]{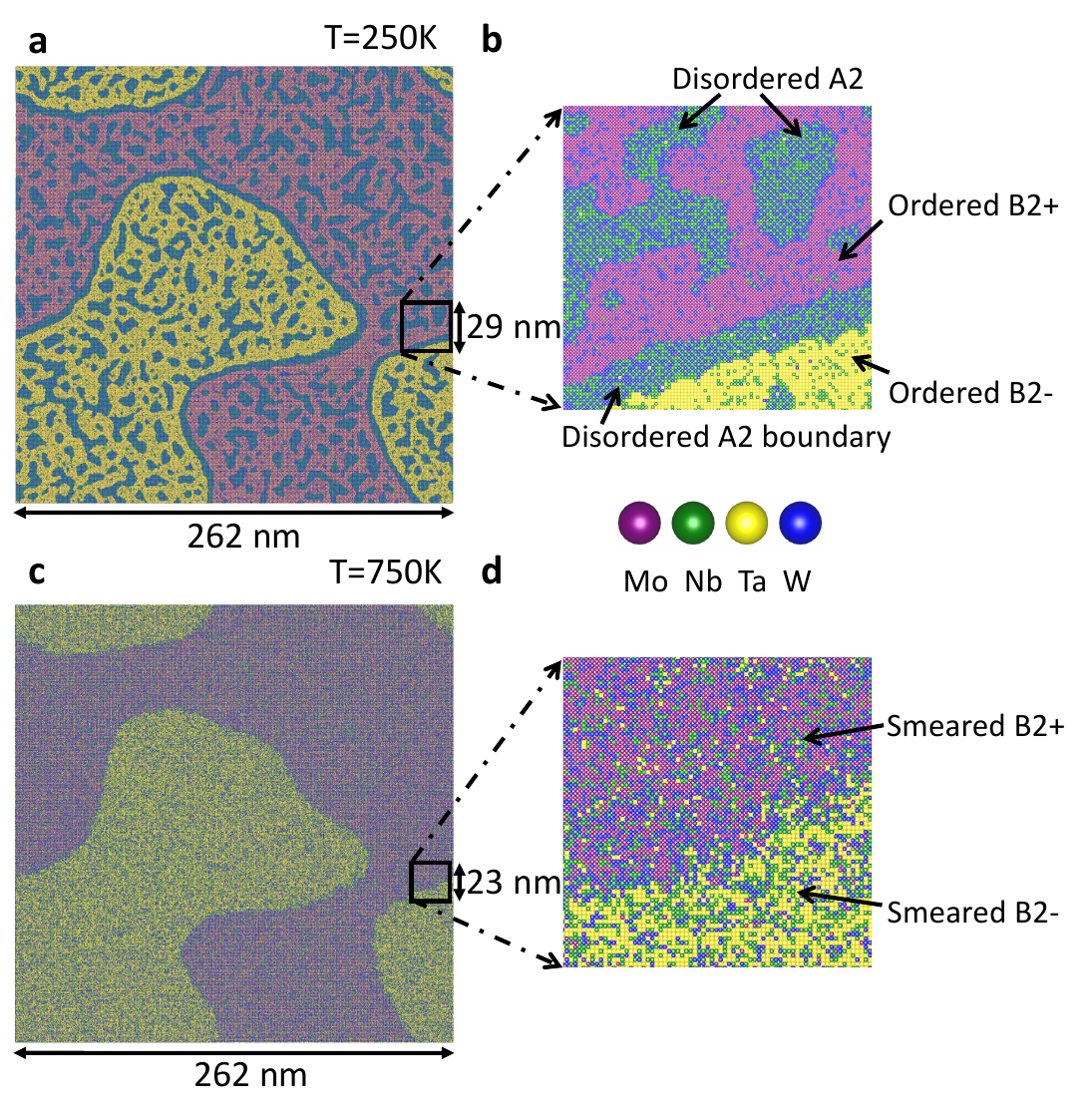}
    \caption{Simulation results for MoNbTaW with 1 billion atoms. (a) Snapshot of the [001] face at T=250 K. (b)  A magnified view of the 29 nm square in a. (c) Snapshot of the [001] face at T=750 K. (d) A magnified view of the 23 nm square in c.}
    \label{fig:MoNbTaW_1B}
\end{figure}
Other than FeCoNiAlTi, we also employ our method to study the nanostructure evolution of a well-studied bcc HEA: MoNbTaW. The size of the supercell is $795\times795\times795\times2 = 1,004,919,750$ atoms. The size of the link-cell is $3\times3\times3$, and the energy model is 2nd-NN EPI. The simulation temperatures decrease from 2000 K to 250K, with a temperature interval of 50 K. For each temperature, we run the simulation for $10^5$ sweeps.  
The simulation results are shown in Fig.~\ref{fig:MoNbTaW_1B}. For T=250K, we note that there are two types of nanostructure: The smaller ones are mainly made up of Nb and W, and of a feature size of about 10 nm. It can be seen that the distributions of Nb and W elements are relatively random. The shapes of these small nanoparticles is also more diverse than that in the FeCoNiAlTi. Other than the small nanostructure, there are also larger nanophases of feature sizes of about 100 nm, which are at the same length scale as the supercell. The two nanophases are shown as yellow and purple in the figure. It is easy to see that they are actually the MoTa B2 phases, which is in agreement with previous results \cite{Korman_npj, LIU2021110135, YAO2024120457}. The difference between the yellow and the purple nanophases is that one of them has Ta at the center position of the perfect bcc lattice (denote as B2- in Fig.~\ref{fig:MoNbTaW_1B}), while the other has Mo at the center site (denote as B2+ in Fig.~\ref{fig:MoNbTaW_1B}). These two nonophases can be understood as the spontaneous symmetry breaking of an Ising-like model along either the plus or the minus directions. At T=250 K, a chemical grain boundary made up of Nb and Ta can be clearly seen between the two nanophases, and the thickness of the chemical grain boundary is about 3 nm. As the temperature increases to 750 K, it can be seen that the larger nanophases still exist, which means that those features are more stable against thermal fluctuations. On the other hand, the smaller nanoparticles vanish, along with the chemical grain boundary between the yellow and purple nanophases. The stability of the nanophases presents an interesting phenomenon, which we propose to explain as follows: at low temperatures, the disordered A2 boundary acts as a protective barrier, preventing the formation of energetically unfavorable AA or BB nearest-neighbor pairs when the two nanophases come into contact. On the other hand, as the temperature increases, the disordered chemical grain boundary will dissolve into the B2 phases, 
which introduces additional chemical disorder. This disorder reduces the energy cost at the boundary, allowing the nanostructures to persist even in the absence of the protective chemical grain boundary. It would be intriguing if this ``disorder protected nanophase" revealed in the simulations could be directly verified by future experiments.

{\color{black}
\subsection{Discussion of limits}
To facilitate better interpretation of the simulation results, we examine the method’s limitations and suggest directions for future refinement. Note that the interaction range used in this work is limited to the first and second nearest neighbors. In contrast, Ref.~\cite{Körmann02012017} suggests that effective pair interactions (EPIs) can be long-ranged, with contributions from up to 99 coordination shells ($p_\text{max}$) required for accurate total energy estimation. However, those EPIs were derived using the EMTO+CPA electronic structure method and are not necessarily equivalent to the EPIs used in this work, which were obtained by regression from direct DFT calculations. In a previous study, Liu \textit{et al.} investigated the influence of interaction range on model performance. Their findings indicate that the optimal cutoff depends on the training dataset, governed by the well-known bias--variance trade-off. In general, the first two coordination shells capture the most significant contributions to the energy, while extending the range can improve accuracy if supported by a larger DFT dataset. Comparing this work, which uses a cutoff of $p_\text{max} = 2$, to Ref.~\cite{Körmann02012017}, which used $p_\text{max} = 6$, the resulting order--disorder transition temperatures are 1050~K and 873~K, respectively. This relatively modest difference supports the adequacy of a short-range model for the current study, given the size of the available DFT dataset. 

A further limitation of this work is that lattice distortions are not included in the DFT calculations, which can potentially change the order-disorder transition temperature significantly, as noticed for MoNbTaW \cite{Korman_npj}. The lack of lattice relaxation in our DFT data stems from the use of the LSMS method, which, while computationally efficient, does not currently support lattice relaxations. Although the distortions in the HEAs studied here are generally small, their precise impact should be addressed in future work. 

Finally, beyond lattice distortion, several other factors can influence phase transitions. A rigorous treatment of these effects requires evaluating all relevant contributions to the Gibbs free energy, including electronic, vibrational (phonon), and configurational terms. When the crystal structure remains unchanged, it is generally reasonable to approximate phase stability using only the configurational term, as chemical disorder predominantly introduces smearing of the Fermi surface and phonon dispersion. Moreover, volume thermal expansion is typically less than 1\% between room temperature and 1000 K \cite{LAPLANCHE2018244}, and such a minor change alone is unlikely to significantly affect the phase transition behavior. Nonetheless, a comprehensive evaluation of all relevant contributions is required for a definitive conclusion, which is beyond the scope of this study.
}
\section{Method}
\subsection{Theoretical speedup of SMC-X vstraditional methods
}
\label{Method:scaling}
The theoretical speedup of the SMC-X method as compared to other methods are estimated as follows: For DFT and linear-DFT, we assume that a 100-atoms SCF calculation takes one hour, based on our experience with the MuST-KKR code \cite{LIU2018265}. The speed of the GNN method (Allegro) is estimated from the $\rm{Li_3PO_4}$ structure of 421,824 atoms in Ref.~\cite{MPNNs_2023}. Note that an MC sweep is defined as making an MC trial move over each lattice site. Combining it with the intrinsic $O(N^3)$ scaling in the DFT method, the total computational cost of DFT then scales as $O(N^4)$, as signified in Fig.~\ref{fig:link-cell}
(d). Building on the principles of near-sightedness \cite{doi:10.1073/pnas.0505436102}, the LSMS method employs an approximation that confines electron scattering to a local interaction zone. This approximation reduces the computational cost of energy evaluation to linear scaling, therefore the total computational cost for an MC sweep scales approximately as $O(N^2)$. 
For a one million-atom system, this reduces the computational cost by a factor of $10^8$, as shown in the orange line in Fig.~\ref{fig:link-cell} (d). Despite the enhanced scaling behavior, linear-DFT still requires solving the Schr\"odinger equations explicitly, which contributes a large prefactor to the total computational cost. By replacing the computationally expensive DFT method with machine learning models such as the GNN model, the computational cost can be further reduced by a factor of $10^8$, as shown in the green line in Fig.~\ref{fig:link-cell} (d). The speedup from the SMC-GPU implementation can be divided into multiple parts: First, the SMC-X parallelization strategy replaces the energy evaluation of the whole system with calculating the energy changes of the LIZ, which is independent of the system size. This strategy reduces the computational complexity from $O(N^2)$ to $O(N)$, as illustrated in Fig.~\ref{fig:link-cell} (d). Second, the SMC-X method can simultaneously harness the thousands or tens of thousands of cores in a modern high-performance GPU (e.g. 14,592 FP32 CUDA cores in an NVIDIA H800 GPU). Third, the energy models used in this work are relatively simple compared to the GNN models. Based on the preceding discussion, the total acceleration ratio of the SMC-GPU method relative to a CPU-based DFT method, for a system of $N$ atoms, can be decomposed into contributions from nearsightedness $S_{NS}$, ML acceleration $S_{ML}$,  and SMC-X implementation $S_{SMC-GPU}$, and written as:
\begin{align}
    S = S_{NS} \times S_{ML} \times S_{SMC-GPU} \approx 10^7\times N^3, \label{speedup}
\end{align}
as illustrated in Fig.~\ref{fig:link-cell} (d).

\subsection{Implementation of SMC-GPU}
\label{Method:SMC-GPU}
For a simple nearest-neighbor 2D Ising model, a widely used parallelization scheme is the checkerboard algorithm. In the checkerboard algorithm, the 2D square lattice is divided into two sublattices, one colored black, and the other white. Note that the lattice sites of the same color will not interact with each other; therefore, the sites of one color can be updated simultaneously by fixing the sites of the other color. The checkerboard algorithm is very efficient and has been applied to study the 2D and 3D Ising models on different accelerators, including GPU \cite{PREIS20094468,BLOCK20101549, LULLI2015290}, TPU \cite{10.1145/3295500.3356149}, and FPGA \cite{LIN2013224, 10.1109/TPDS.2015.2505725}. 

The checkerboard algorithm cannot be directly applied when the Hamiltonian contains interactions beyond the nearest neighbors.
Therefore, it is rarely used for studying real materials, where the interaction range is typically beyond nearest neighbors.
For such a purpose, a generalization of the checkerboard algorithm can be employed to harness the move-parallelism opportunity. This method is referred to as a parallelized link-cell algorithm \cite{Yamakov2016ParallelGC}. 
In the link-cell algorithm, the lattice is decomposed into $n_C$ cells of fixed size $a_{cell}$, which should be larger than the interaction range. With the link-cell algorithm, the serial MC code can be accelerated up to $n_C$ times. 
For a multi-GPU system, the $n_C$ cells are evenly distributed on all the GPUs, and ghost boundaries are added to take into account the interactions between atoms from different GPUs, as shown in Fig.~\ref{fig:link-cell} (b).

\begin{figure} [ht!]
    \centering
    \includegraphics[width=0.7 \linewidth]{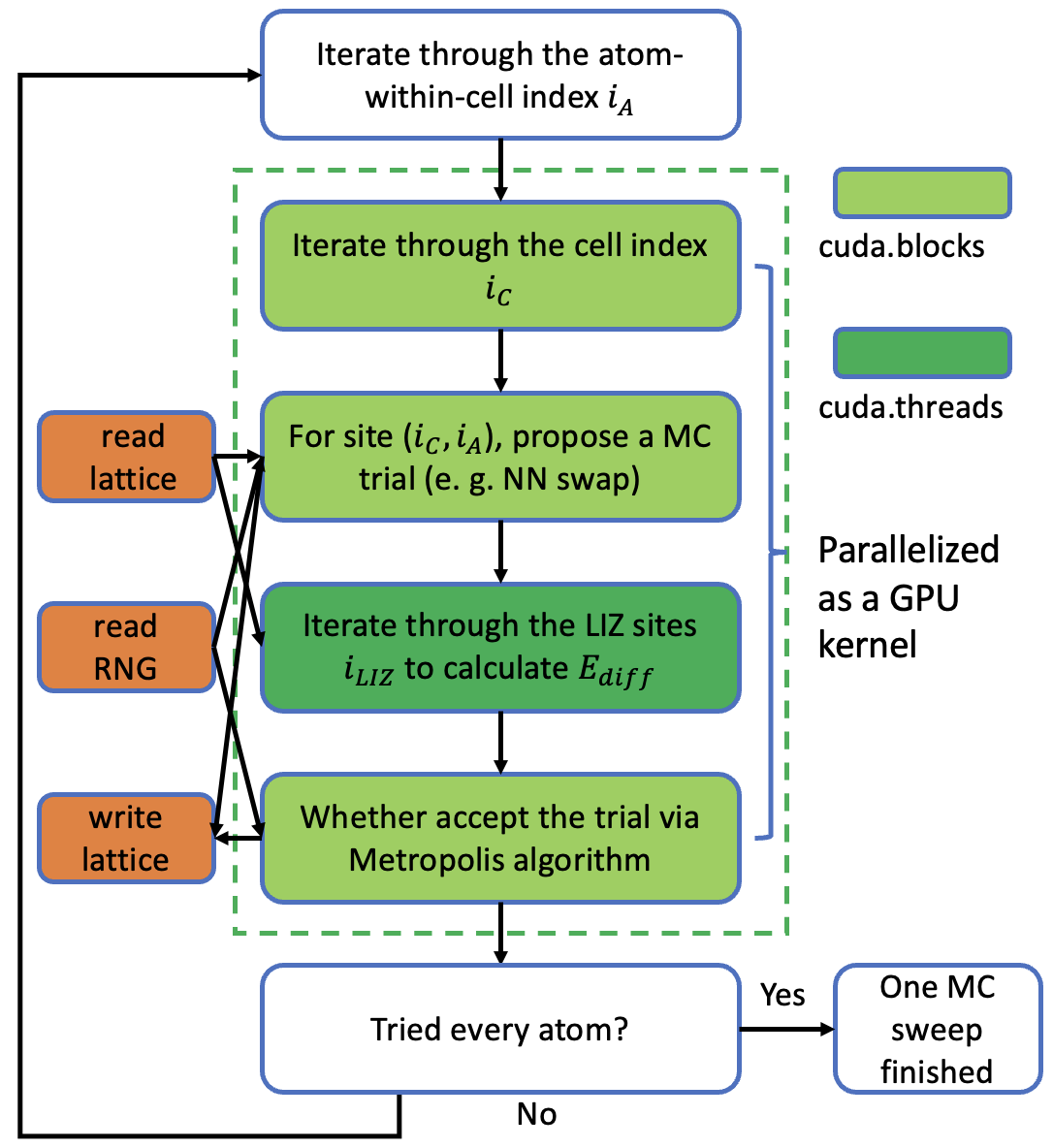}
    \caption{A schematic flowchart for the link-cell parallelism algorithm for a machine learning energy model. The outer loop over index $i_A$ is executed sequentially, while the inner loop over index $i_C$ is executed in parallel as a CUDA kernel since the MC update atoms of the same $i_A$ index but different $i_C$ indices are independent. }
    \label{fig:flowchart}
\end{figure}

\begin{itemize}
    \item Parallelization implementation: The actual parallelization implementation on GPUs depends on the energy model. For pairwise models such as EPI, the iteration over the $i_C$ index is executed in parallel as CUDA threads on the GPU. For more general energy models such as the qSRO model, the $i_C$ index is parallelized in the CUDA blocks dimension, and the calculation of the local energies for each site within LIZ is parallelized with the CUDA threads, as demonstrated in the flowchart in Fig.~\ref{fig:flowchart}.
    \item Link-cell size: Using the link-cell algorithm, the one MC sweep (MC trials over every sites) is decomposed into $n_A$ sequential steps, with each one of $n_C$ parallel moves. For instance, for a $300\times300\times300$ bcc supercell (a total of $300^3\times2=54,000,000$ atoms), assuming next-to-nearest-neighbor EPI interaction, then the link-cell length can be chosen as $3\times a$, where a is the lattice constant. The link-cell size $n_A$ is $3\times3\times3\times2=54$, and the number of link-cell is $n_C=1,000,000$.
    \item Random number generator: An efficient random number generator is important for MC simulation because random numbers are needed for every MC trial, as shown in the fourth and fifth steps in Fig.~\ref{fig:flowchart}. Our implementation makes use of the NVIDIA CUDA Random Number Generation library (cuRAND), which delivers high performance GPU-accelerated pseudorandom numbers. 
    \item Search neighbors: Another important consideration is how to efficiently find the neighboring sites, for which their chemical species need to be read from the memory, as shown in the third step in Fig.~\ref{fig:flowchart}. In practice, we make use of the lattice structure and directly calculate the indices of the neighbors, as well as the features of the local chemical environment, on-the-fly using a list of the relative positions of the neighboring atoms. This method has a constant time complexity, which is advantageous over the more common practice of using k-D tree to store the atomic positions and search the neighbors, which has a time complexity of $O(\log(N))$. 
\end{itemize}

\subsection{The EPI and qSRO model}
\label{Method:model}
In the EPI model, the effective Hamiltonian of the system is made up of chemical pair interactions of the centering atom with neighboring atoms within some cutoff radius. The local chemical environment is specified by $\vec{\sigma} = (\sigma^0, \sigma^1, \cdots, \sigma^{N_{n}-1})$, which denotes the chemical species of the $N_n$-th neighboring atoms. The local energy $E_i$ is given by 
\begin{align}
E_i = \sum_{f} V^f \pi^f(\vec{\sigma}_i)  + V^p_i + V^0 + \epsilon,
\label{Linear_EPI}
\end{align}
where $\epsilon$ is the uncertainty of the EPI model, $V^0$ is the bias term same for all sites, $V_i^p$ is a single-site term depending only on the chemical component $p$ of atom $i$, $V^f$ are the EPI parameters, and $\pi^f$ are the number of pair interactions of type $f$. The feature index $f$ is actually made up of three parts $(p, p', m)$, representing the element of the local atom, the element of the neighboring atoms, and the coordination shell, respectively. For a system of fixed chemical concentrations, summing up the local energies over all sites, the total energy is then given by
\begin{align}
    {\color{black}E = N \sum_{p'<p,m} V_m^{pp'} \Pi_m^{pp'} + \mathrm{const} + \epsilon,}
\label{Total_EPI}
\end{align}
where $N$ is the total number of atoms and $\Pi_{m}^{pp'}$ is the proportion of $pp'$ interaction in the $m$-th neighboring shell. Note that due to the fixed chemical concentration in the canonical system, the single-site term $V_i^p$ has been absorbed into the constant, and the number of independent EPI parameters is $M(M-1)/2$ for a $M$-component system, which is the reason for the $p'<p$ requirement. In practice, the EPI parameters are determined via Bayesian regression \cite{LIU2021110135} or stochastic gradient descent. 

As mentioned, since the EPI model contains only pairwise interactions, the calculation of the energy changes after an MC swap trial is simple: we simply multiply the local energy changes of the two swapping sites by a factor of 2 to take into account the local energy changes of other sites. For more general energy models that contain higher-order/nonlinear interactions, the total energy changes need to be explicitly calculated by adding up the local energies of all sites within the LIZ. As a demonstration of our method for more general machine learning models, here we modify the EPI model by adding a simple quadratic term, and the local energy for each site $E_i$ is:
\begin{align}
    {\color{black}E_i = \sum_{p'<p,m} V_m^{pp'} \pi_m^{pp'} + \sum_{p,m} W_m^p (\pi_m^{pp})^2 + \lambda \sum_{p,m} (W_m^p)^2+\mathrm{const} + \epsilon,}
\label{quadratic_local_SRO}
\end{align}
in which 
\begin{align}
    \pi_m^{pp'} = \frac{n_m^{pp'}}{\sum_{pp'} n_m^{pp'}}
\end{align}
is the percentage of $pp'$ interactions for the $m$'s coordination shell, and $n_m^{pp'}$ is the number of pairs with index $(m,p,p')$. Note that we limit the quadratic interaction to the same-element pairs to reduce the number of higher-order terms. The third term on the right of Eq.~\ref{quadratic_local_SRO} is a $l^2$ regularization of the weights of the quadratic term, reflecting our prior that the higher-order interactions should be a small correction to the EPI model. We name it the qSRO model since $\Pi_m^{pp'}$ describes the short-range order within the local chemical environment. It should be noted that, unlike EPI, this qSRO model is no longer a linear or even quadratic function of the pairwise interactions. Instead, it can be expressed as a power series (infinite degree polynomial) of the pairwise interactions due to the $(\pi_m^{pp'})^2$ term. It is easy to see that the nonlinear interactions can also be modeled with neural networks such as the multilayer perceptrons:
\begin{align}
    {\color{black}E_i = \sum_{p'<p,m} V_m^{pp'} \pi_m^{pp'} + \rm{MLP}(\pi_m^{pp'}) + \epsilon.}
\label{quadratic_local_SRO}
\end{align}
However, due to a large number of principal elements, the number of parameters in such an MLP model can easily go beyond the number of data points used in this work (e.g. 220 for FeCoNiAlTi), which leads to a high risk of overfitting. Therefore, in this work we will use the local SRO model as a representation of general ML energy models, and leave more complex models for future study when significantly larger DFT datasets are available.  

\subsection{DFT dataset}
\label{Method:DFT_data}
We use the LSMS method \cite{PhysRevLett.75.2867} to calculate the total energy of the system. LSMS is an all-electron electronic structure calculation method, in which the computational cost scales linearly with respect to the number of atoms. For FeCoNiAlTi, we use a 100-atom supercell with a lattice constant of 6.72 Bohr (0.356 nm). We employs a spin-polarized scheme to account for the magnetic interactions in the system. The angular momentum cutoff $l_{max}$ for the electron wavefunctions is chosen as 3, and the LIZ is chosen as 86 atoms. We used PBE as the exchange-correlational functional. To enhance the representativeness of the DFT data, the total dataset is made up of 120 random generated configurations and 100 configurations from Monte Carlo simulation, as proposed in Ref.~\cite{LIU2021110135}. Therefore the total number of configurations is 220, with each one comprising 100 atoms. 

For MoNbTaW, the DFT dataset has been reported in Ref.~\cite{LIU2021110135}, which is also made up of the random samples and the MC samples. The 704 random samples are calculated with supercells of 64, 128, 256, and 512 atoms. The 72 MC samples are obtained from MC simulations at different temperatures, on a 1000-atom supercell. When calculate the energies of the MoNbTaW system, the lattice constant is set at 6.2 Bohr, and the angular momentum cutoff is chosen as 3. The Barth-Hedin local-density approximation are used as the exchange-correlation functional. The local interaction zone is maintained at 59 atoms. To adequately capture the effects of heavier elements in the system, the scalar-relativistic equation is used instead of the conventional Schr\"{o}dinger equation.

\subsection{MC simulation}
\label{Method:MC}
\begin{table}[htbp]
\centering
\begin{tabular}{|l|r|r|c|c|c|}
\hline
\multicolumn{1}{|c|}{System\_Model} & \multicolumn{1}{c|}{Dimensions} & \multicolumn{1}{c|}{Link-cell} & \multicolumn{1}{c|}{$\rm{T_{sweep}(s)}$} & \multicolumn{1}{c|}{$\rm{N_{sweeps}}/T$} \\
\hline
FeCoNiAlTi\_1M\_EPI & $72 \times 60 \times 60 \times 4 $ & $3\times3\times3$ & 0.001755 & $1.5\times10^6$ (1000\;K) \\
FeCoNiAlTi\_1M\_qSRO & $72 \times 60 \times 60 \times 4 $ & $\rm{4\times 4 \times 4}$ & 0.04703 & $10^5$ (1000\;K) \\
FeCoNiAlTi\_1B\_EPI & $630 \times 630 \times 630 \times 4 $ & $\rm{3\times3\times3}$ & 0.4120 & $10^5$ (1000\;K) \\
FeCoNiAlTi\_1B\_qSRO & $632 \times 632 \times 632 \times 4 $ & $\rm{4\times4\times4}$ & 21.65 & $10^4$ (1000\;K) \\
MoNbTaW\_1B\_EPI & $795 \times 795 \times 795 \times 2$ & $\rm{3\times3\times3}$ & 0.1912 & $10^5$ (50\;K) \\
\hline
\end{tabular}
\caption{The systems and models used for the MC simulation. The dimensions are made up of $nx\times ny\times nz \times m$, where $nx, ny, nz$ are the supercell vectors along the $x, y, z$ direction, and $m$ is the number of atoms in the cubic cell. The times of one MC sweep $\rm{T_{sweep}}$ for different cases are also shown to evaluate the speed.}
\label{tab:models_summary}
\end{table}
The results shown in this work are based on five simulation results using different models and systems, as shown in Tab.~\ref{tab:models_summary}. The MC simulation is a simulated annealing process in the canonical ensemble (NVT). The temperature is initialized as 2000 K and then decrease with a step of 50 K. The initial $\rm{N_{sweeps}}$ sweeps
are discarded before making measurement or taking a snapshot of the configuration. The number of $\rm{N_{sweeps}}$ generally increases as the temperature decreases to account for the lower acceptance ratio of MC moves. For instance, for $\rm{FeCoNiAlTi\_1M\_EPI}$, the number of sweeps is $4\times 10^5$ at 2000 K, then increase to $1.5\times 10^6$ at $T=1000 K$. For $\rm{FeCoNiAlTi\_1B\_EPI}$, we evenly decrease the simulation temperature from 2000 K to 1000 K using a step of 50 K. For the 1B atom system, we no longer record the MC configurations and run $10^5$ MC sweeps before recording the configuration at $T=1000$ K. Note that, generally speaking, the $10^5$ number of MC sweeps is not large enough for a supercell as large as one billion atoms to reach thermal equilibrium, so the simulation is closer to simulated annealing. Nevertheless, this nonequilibrium does not necessarily present as a problem since the actual synthesis of alloys is typically a nonequilibrium process involving a rapid heating-cooling process, in which the phases at high-temperature can be trapped by the kinetic barrier and be maintained at low temperatures. 

In order to extract the nanostructures from the simulation results, we employs the union-find algorithm, which uses a tree structure to efficiently manage and manipulate partitions of a set into disjoint subsets. In order to apply the union-find algorithm, we first calculate the local energy of each site and set a threshold of the energy to determine whether two sites belong to the same set (cluster). MC simulations are performed on a workstation with two 80 GB NVIDIA H800 SXM5 GPUs that are connected via 400 GB/s bidirectional bandwidth NVLink. In practice, we find that one GPU is large enough to fit the one-billion atom system, and the two GPUs are simultaneously used via either temperature parallelization or lattice decomposition, as described in section.~\ref{Results:SMC}. For visualization, we used VESTA \cite{VESTA} to generate the 3D crystals and nanoparticles figures.

\section{Conclusions}
We introduce a Scalable Monte Carlo (SMC-X) method that overcomes the parallelization bottlenecks inherent in conventional Monte Carlo (MC) simulations. By generalizing the checkerboard algorithm through the introduction of link cells, our approach reduces the computational complexity of an MC sweep from $O(N^2)$ to $O(N)$. This method is not only applicable to pairwise interactions but also extends to nonlinear local interactions through the introduction of a local interaction zone, making it highly compatible with machine learning (ML)-enhanced atomistic models.

The GPU-accelerated implementation of the SMC-X method enables the simulation of atomistic systems exceeding one billion atoms while maintaining the accuracy of density functional theory (DFT). Such unprecedented capability makes it possible for us to directly observe the nanostructures in HEAs, which can be of more than millions of atoms. Understanding the nanoscale evolution of NPs are vital for understanding the origin of the superb mechanical properties in HEAs. 
    
Using the SMC-GPU code, we investigated the size, composition, and morphologies of the NPs in the fcc $\rm{Fe_{29}Co_{29}Ni_{28}Al_7Ti_7}$ HEA using a one-billion-atom supercell. This large supercell not only makes it possible to directly observe the nanostructure evolution in HEAs, but also enables the direct comparison with experimental results from TEM and APT. This size and composition of the nanoparticles generally align well with available experimental findings \cite{Yang933}. Moreover, we find that seemingly separate NPs may, in fact, be connected, highlighting the intricate nature of high-entropy alloys (HEAs) and prompting a reconsideration of traditional grain size measurement methods. 
    
We further investigated the bcc MoNbTaW high-entropy alloy (HEA) using a one-billion-atom supercell. The results reveal the formation of hierarchical nanostructures: The smaller NPs consist of a disordered A2 structure comprising Nb and W, with a feature size of approximately 10 nm; The larger nanophases exhibits a B2 structure composed of Ta and Mo, with a feature size of about 100 nm. Interestingly, we find that the Ta-Mo B2 structures decompose into two distinct nanophases, B2+ and B2-, which are separated by a grain boundary enriched with Nb and W. The simulation results suggest that the disordered Nb-W grain boundary acts as a protective barrier for the two nanophases to reduce the enthalpy of the system.

{\color{black}
\section*{Code availability statement}
Executable binaries or a containerized version of the code are available from the authors upon reasonable request.
}

\section*{Data availability statement}
The data supporting the findings of this study, including the simulated configurations in .xyz format, Python scripts for data analysis, andC++ code for nanoparticle extraction, are available at:
https://github.com/xianglil/SMCX-Supplementary

\section*{Acknowledgements}
The work of X. Liu and F. Zhou was supported by the National Natural Science Foundation of China under Grant 12404283. The work of Y. Tian was supported by the National Natural Science Foundation of China under Grant Grant 62425101, 62088102. This work was also supported by the Pengcheng Laboratory Key Project (PCL2021A13) to utilized the computing resources of Pengcheng Cloud Brain. 

{\color{black}
X. Liu gratefully acknowledges Professor Robert Swendsen for his guidance and foundational training in the fascinating field of Monte Carlo simulation during his graduate studies at Carnegie Mellon University. 
}
\section*{Author contributions}
X.L. proposed the SMC method, designed the workflow, calculated the DFT data, trained the energy model, interpreted the data, acquiredthe funding, supervised the work, and wrote the main manuscript text. K.Y. wrote the SMC-GPU code based on X.L.'s prototype, and ranthe MC simulations. F.Z. contributed to Fig.~1d
, 4a, 4b and Table 1. Z.P. co-analyzed the data. P.X. co-analyzed the data, acquired the computing resources, and co-supervised the work. Y.T. acquired the funding. All authors reviewed the manuscript.

\subsection*{Corresponding authors}
Correspondence to Xianglin Liu (submitting) (xianglinliu01@gmail.com) or Pengxiang Xu (xupx@pcl.ac.cn)

\bibliographystyle{model1-num-names}
\bibliography{sample.bib}

\begin{thebibliography}{93}
\expandafter\ifx\csname natexlab\endcsname\relax\def\natexlab#1{#1}\fi
\providecommand{\bibinfo}[2]{#2}
\ifx\xfnm\relax \def\xfnm[#1]{\unskip,\space#1}\fi
\bibitem[{Burke(2012)}]{10.1063/1.4704546}
\bibinfo{author}{K.~Burke},
\newblock \bibinfo{title}{{Perspective on density functional theory}},
\newblock \bibinfo{journal}{The Journal of Chemical Physics} \bibinfo{volume}{136} (\bibinfo{year}{2012}). \bibinfo{note}{150901}.
\bibitem[{Ma et~al.(2015)Ma, Grabowski, Körmann, Neugebauer, and Raabe}]{MA201590}
\bibinfo{author}{D.~Ma}, \bibinfo{author}{B.~Grabowski}, \bibinfo{author}{F.~Körmann}, \bibinfo{author}{J.~Neugebauer}, \bibinfo{author}{D.~Raabe},
\newblock \bibinfo{title}{Ab initio thermodynamics of the cocrfemnni high entropy alloy: Importance of entropy contributions beyond the configurational one},
\newblock \bibinfo{journal}{Acta Materialia} \bibinfo{volume}{100} (\bibinfo{year}{2015}) \bibinfo{pages}{90--97}.
\bibitem[{Khan and Eisenbach(2016)}]{PhysRevB.93.024203}
\bibinfo{author}{S.~N. Khan}, \bibinfo{author}{M.~Eisenbach},
\newblock \bibinfo{title}{{Density-functional Monte-Carlo simulation of CuZn order-disorder transition}},
\newblock \bibinfo{journal}{Phys. Rev. B} \bibinfo{volume}{93} (\bibinfo{year}{2016}) \bibinfo{pages}{024203}.
\bibitem[{Wilson(1975)}]{RevModPhys.47.773}
\bibinfo{author}{K.~G. Wilson},
\newblock \bibinfo{title}{The renormalization group: Critical phenomena and the kondo problem},
\newblock \bibinfo{journal}{Rev. Mod. Phys.} \bibinfo{volume}{47} (\bibinfo{year}{1975}) \bibinfo{pages}{773--840}.
\bibitem[{Sanchez(2019)}]{PhysRevB.99.134206}
\bibinfo{author}{J.~M. Sanchez},
\newblock \bibinfo{title}{Renormalized interactions in truncated cluster expansions},
\newblock \bibinfo{journal}{Phys. Rev. B} \bibinfo{volume}{99} (\bibinfo{year}{2019}) \bibinfo{pages}{134206}.
\bibitem[{Pawley et~al.(1984)Pawley, Swendsen, Wallace, and Wilson}]{PhysRevB.29.4030}
\bibinfo{author}{G.~S. Pawley}, \bibinfo{author}{R.~H. Swendsen}, \bibinfo{author}{D.~J. Wallace}, \bibinfo{author}{K.~G. Wilson},
\newblock \bibinfo{title}{Monte carlo renormalization-group calculations of critical behavior in the simple-cubic ising model},
\newblock \bibinfo{journal}{Phys. Rev. B} \bibinfo{volume}{29} (\bibinfo{year}{1984}) \bibinfo{pages}{4030--4040}.
\bibitem[{Park et~al.(2021)Park, Kornbluth, Vandermause, Wolverton, Kozinsky, and Mailoa}]{GNNFF_NPJ_2021}
\bibinfo{author}{C.~W. Park}, \bibinfo{author}{M.~Kornbluth}, \bibinfo{author}{J.~Vandermause}, \bibinfo{author}{C.~Wolverton}, \bibinfo{author}{B.~Kozinsky}, \bibinfo{author}{J.~P. Mailoa},
\newblock \bibinfo{title}{Accurate and scalable graph neural network force field and molecular dynamics with direct force architecture},
\newblock \bibinfo{journal}{npj Computational Materials} \bibinfo{volume}{7} (\bibinfo{year}{2021}) \bibinfo{pages}{73}.
\bibitem[{Batzner et~al.(2022)Batzner, Musaelian, Sun, Geiger, Mailoa, Kornbluth, Molinari, Smidt, and Kozinsky}]{NequIP_NC}
\bibinfo{author}{S.~Batzner}, \bibinfo{author}{A.~Musaelian}, \bibinfo{author}{L.~Sun}, \bibinfo{author}{M.~Geiger}, \bibinfo{author}{J.~P. Mailoa}, \bibinfo{author}{M.~Kornbluth}, \bibinfo{author}{N.~Molinari}, \bibinfo{author}{T.~E. Smidt}, \bibinfo{author}{B.~Kozinsky},
\newblock \bibinfo{title}{E(3)-equivariant graph neural networks for data-efficient and accurate interatomic potentials},
\newblock \bibinfo{journal}{Nature Communications} \bibinfo{volume}{13} (\bibinfo{year}{2022}) \bibinfo{pages}{2453}.
\bibitem[{Song et~al.(2024)Song, Zhao, Liu, Wang, Lindgren, Wang, Chen, Xu, Liang, Ying, Xu, Zhao, Shi, Wang, Lyu, Zeng, Liang, Dong, Sun, Chen, Zhang, Guo, Qian, Sun, Erhart, Ala-Nissila, Su, and Fan}]{MLPAlloys2024}
\bibinfo{author}{K.~Song}, \bibinfo{author}{R.~Zhao}, \bibinfo{author}{J.~Liu}, \bibinfo{author}{Y.~Wang}, \bibinfo{author}{E.~Lindgren}, \bibinfo{author}{Y.~Wang}, \bibinfo{author}{S.~Chen}, \bibinfo{author}{K.~Xu}, \bibinfo{author}{T.~Liang}, \bibinfo{author}{P.~Ying}, \bibinfo{author}{N.~Xu}, \bibinfo{author}{Z.~Zhao}, \bibinfo{author}{J.~Shi}, \bibinfo{author}{J.~Wang}, \bibinfo{author}{S.~Lyu}, \bibinfo{author}{Z.~Zeng}, \bibinfo{author}{S.~Liang}, \bibinfo{author}{H.~Dong}, \bibinfo{author}{L.~Sun}, \bibinfo{author}{Y.~Chen}, \bibinfo{author}{Z.~Zhang}, \bibinfo{author}{W.~Guo}, \bibinfo{author}{P.~Qian}, \bibinfo{author}{J.~Sun}, \bibinfo{author}{P.~Erhart}, \bibinfo{author}{T.~Ala-Nissila}, \bibinfo{author}{Y.~Su}, \bibinfo{author}{Z.~Fan},
\newblock \bibinfo{title}{General-purpose machine-learned potential for 16 elemental metals and their alloys},
\newblock \bibinfo{journal}{Nature Communications} \bibinfo{volume}{15} (\bibinfo{year}{2024}) \bibinfo{pages}{10208}.
\bibitem[{Zeni et~al.(2025)Zeni, Pinsler, Z{\"u}gner, Fowler, Horton, Fu, Wang, Shysheya, Crabb{\'e}, Ueda, Sordillo, Sun, Smith, Nguyen, Schulz, Lewis, Huang, Lu, Zhou, Yang, Hao, Li, Yang, Li, Tomioka, and Xie}]{2025Nature}
\bibinfo{author}{C.~Zeni}, \bibinfo{author}{R.~Pinsler}, \bibinfo{author}{D.~Z{\"u}gner}, \bibinfo{author}{A.~Fowler}, \bibinfo{author}{M.~Horton}, \bibinfo{author}{X.~Fu}, \bibinfo{author}{Z.~Wang}, \bibinfo{author}{A.~Shysheya}, \bibinfo{author}{J.~Crabb{\'e}}, \bibinfo{author}{S.~Ueda}, \bibinfo{author}{R.~Sordillo}, \bibinfo{author}{L.~Sun}, \bibinfo{author}{J.~Smith}, \bibinfo{author}{B.~Nguyen}, \bibinfo{author}{H.~Schulz}, \bibinfo{author}{S.~Lewis}, \bibinfo{author}{C.-W. Huang}, \bibinfo{author}{Z.~Lu}, \bibinfo{author}{Y.~Zhou}, \bibinfo{author}{H.~Yang}, \bibinfo{author}{H.~Hao}, \bibinfo{author}{J.~Li}, \bibinfo{author}{C.~Yang}, \bibinfo{author}{W.~Li}, \bibinfo{author}{R.~Tomioka}, \bibinfo{author}{T.~Xie},
\newblock \bibinfo{title}{A generative model for inorganic materials design},
\newblock \bibinfo{journal}{Nature}  (\bibinfo{year}{2025}).
\bibitem[{Ko et~al.(2023)Ko, Finkler, Goedecker, and Behler}]{doi:10.1021/acs.jctc.2c01146}
\bibinfo{author}{T.~W. Ko}, \bibinfo{author}{J.~A. Finkler}, \bibinfo{author}{S.~Goedecker}, \bibinfo{author}{J.~Behler},
\newblock \bibinfo{title}{Accurate fourth-generation machine learning potentials by electrostatic embedding},
\newblock \bibinfo{journal}{Journal of Chemical Theory and Computation} \bibinfo{volume}{19} (\bibinfo{year}{2023}) \bibinfo{pages}{3567--3579}. \bibinfo{note}{PMID: 37289440}.
\bibitem[{Bart\'ok et~al.(2010)Bart\'ok, Payne, Kondor, and Cs\'anyi}]{PhysRevLett.104.136403}
\bibinfo{author}{A.~P. Bart\'ok}, \bibinfo{author}{M.~C. Payne}, \bibinfo{author}{R.~Kondor}, \bibinfo{author}{G.~Cs\'anyi},
\newblock \bibinfo{title}{Gaussian approximation potentials: The accuracy of quantum mechanics, without the electrons},
\newblock \bibinfo{journal}{Phys. Rev. Lett.} \bibinfo{volume}{104} (\bibinfo{year}{2010}) \bibinfo{pages}{136403}.
\bibitem[{Sch{\"u}tt et~al.(2019)Sch{\"u}tt, Kessel, Gastegger, Nicoli, Tkatchenko, and M{\"u}ller}]{SchnetPack}
\bibinfo{author}{K.~T. Sch{\"u}tt}, \bibinfo{author}{P.~Kessel}, \bibinfo{author}{M.~Gastegger}, \bibinfo{author}{K.~A. Nicoli}, \bibinfo{author}{A.~Tkatchenko}, \bibinfo{author}{K.-R. M{\"u}ller},
\newblock \bibinfo{title}{Schnetpack: A deep learning toolbox for atomistic systems},
\newblock \bibinfo{journal}{Journal of Chemical Theory and Computation} \bibinfo{volume}{15} (\bibinfo{year}{2019}) \bibinfo{pages}{448--455}.
\bibitem[{Gasteiger et~al.(2021)Gasteiger, Becker, and G{\"u}nnemann}]{gasteiger_gemnet_2021}
\bibinfo{author}{J.~Gasteiger}, \bibinfo{author}{F.~Becker}, \bibinfo{author}{S.~G{\"u}nnemann},
\newblock \bibinfo{title}{Gemnet: Universal directional graph neural networks for molecules},
\newblock in: \bibinfo{booktitle}{Conference on Neural Information Processing Systems (NeurIPS)}.
\bibitem[{Ying et~al.(2021)Ying, Cai, Luo, Zheng, Ke, He, Shen, and Liu}]{ying2021do}
\bibinfo{author}{C.~Ying}, \bibinfo{author}{T.~Cai}, \bibinfo{author}{S.~Luo}, \bibinfo{author}{S.~Zheng}, \bibinfo{author}{G.~Ke}, \bibinfo{author}{D.~He}, \bibinfo{author}{Y.~Shen}, \bibinfo{author}{T.-Y. Liu},
\newblock \bibinfo{title}{Do transformers really perform badly for graph representation?},
\newblock in: \bibinfo{booktitle}{Thirty-Fifth Conference on Neural Information Processing Systems}.
\bibitem[{Hart et~al.(2021)Hart, Mueller, Toher, and Curtarolo}]{ML_alloys}
\bibinfo{author}{G.~L.~W. Hart}, \bibinfo{author}{T.~Mueller}, \bibinfo{author}{C.~Toher}, \bibinfo{author}{S.~Curtarolo},
\newblock \bibinfo{title}{Machine learning for alloys},
\newblock \bibinfo{journal}{Nature Reviews Materials} \bibinfo{volume}{6} (\bibinfo{year}{2021}) \bibinfo{pages}{730--755}.
\bibitem[{Liu et~al.(2023)Liu, Zhang, and Pei}]{LIU2023101018}
\bibinfo{author}{X.~Liu}, \bibinfo{author}{J.~Zhang}, \bibinfo{author}{Z.~Pei},
\newblock \bibinfo{title}{Machine learning for high-entropy alloys: Progress, challenges and opportunities},
\newblock \bibinfo{journal}{Progress in Materials Science} \bibinfo{volume}{131} (\bibinfo{year}{2023}) \bibinfo{pages}{101018}.
\bibitem[{Zhang et~al.(2025)Zhang, Sorkin, Aitken, Politano, Behler, P~Thompson, Ko, Ong, Chalykh, Korogod, Podryabinkin, Shapeev, Li, Mishin, Pei, Liu, Kim, Park, Hwang, Han, Sheriff, Cao, and Freitas}]{Zhang_2025}
\bibinfo{author}{Y.-W. Zhang}, \bibinfo{author}{V.~Sorkin}, \bibinfo{author}{Z.~H. Aitken}, \bibinfo{author}{A.~Politano}, \bibinfo{author}{J.~Behler}, \bibinfo{author}{A.~P~Thompson}, \bibinfo{author}{T.~W. Ko}, \bibinfo{author}{S.~P. Ong}, \bibinfo{author}{O.~Chalykh}, \bibinfo{author}{D.~Korogod}, \bibinfo{author}{E.~Podryabinkin}, \bibinfo{author}{A.~Shapeev}, \bibinfo{author}{J.~Li}, \bibinfo{author}{Y.~Mishin}, \bibinfo{author}{Z.~Pei}, \bibinfo{author}{X.~Liu}, \bibinfo{author}{J.~Kim}, \bibinfo{author}{Y.~Park}, \bibinfo{author}{S.~Hwang}, \bibinfo{author}{S.~Han}, \bibinfo{author}{K.~Sheriff}, \bibinfo{author}{Y.~Cao}, \bibinfo{author}{R.~Freitas},
\newblock \bibinfo{title}{Roadmap for the development of machine learning-based interatomic potentials},
\newblock \bibinfo{journal}{Modelling and Simulation in Materials Science and Engineering} \bibinfo{volume}{33} (\bibinfo{year}{2025}) \bibinfo{pages}{023301}.
\bibitem[{Clausen et~al.(2024)Clausen, Rossmeisl, and Ulissi}]{doi:10.1021/acs.jpcc.4c01704}
\bibinfo{author}{C.~M. Clausen}, \bibinfo{author}{J.~Rossmeisl}, \bibinfo{author}{Z.~W. Ulissi},
\newblock \bibinfo{title}{Adapting oc20-trained equiformerv2 models for high-entropy materials},
\newblock \bibinfo{journal}{The Journal of Physical Chemistry C} \bibinfo{volume}{128} (\bibinfo{year}{2024}) \bibinfo{pages}{11190--11195}.
\bibitem[{Chen and Ong(2022)}]{UniversalModel_NCS}
\bibinfo{author}{C.~Chen}, \bibinfo{author}{S.~P. Ong},
\newblock \bibinfo{title}{A universal graph deep learning interatomic potential for the periodic table},
\newblock \bibinfo{journal}{Nature Computational Science} \bibinfo{volume}{2} (\bibinfo{year}{2022}) \bibinfo{pages}{718--728}.
\bibitem[{Huang et~al.(2023)Huang, von Rudorff, and von Lilienfeld}]{doi:10.1126/science.abn3445}
\bibinfo{author}{B.~Huang}, \bibinfo{author}{G.~F. von Rudorff}, \bibinfo{author}{O.~A. von Lilienfeld},
\newblock \bibinfo{title}{The central role of density functional theory in the ai age},
\newblock \bibinfo{journal}{Science} \bibinfo{volume}{381} (\bibinfo{year}{2023}) \bibinfo{pages}{170--175}.
\bibitem[{Jia et~al.(2020)Jia, Wang, Chen, Lu, Lin, Car, E, and Zhang}]{10.5555/3433701.3433707}
\bibinfo{author}{W.~Jia}, \bibinfo{author}{H.~Wang}, \bibinfo{author}{M.~Chen}, \bibinfo{author}{D.~Lu}, \bibinfo{author}{L.~Lin}, \bibinfo{author}{R.~Car}, \bibinfo{author}{W.~E}, \bibinfo{author}{L.~Zhang}, \bibinfo{title}{Pushing the Limit of Molecular Dynamics with Ab Initio Accuracy to 100 Million Atoms with Machine Learning}, \bibinfo{publisher}{IEEE Press}.
\bibitem[{Deringer et~al.(2021)Deringer, Bernstein, Cs{\'a}nyi, Ben~Mahmoud, Ceriotti, Wilson, Drabold, and Elliott}]{Nature_2021_Silicon}
\bibinfo{author}{V.~L. Deringer}, \bibinfo{author}{N.~Bernstein}, \bibinfo{author}{G.~Cs{\'a}nyi}, \bibinfo{author}{C.~Ben~Mahmoud}, \bibinfo{author}{M.~Ceriotti}, \bibinfo{author}{M.~Wilson}, \bibinfo{author}{D.~A. Drabold}, \bibinfo{author}{S.~R. Elliott},
\newblock \bibinfo{title}{Origins of structural and electronic transitions in disordered silicon},
\newblock \bibinfo{journal}{Nature} \bibinfo{volume}{589} (\bibinfo{year}{2021}) \bibinfo{pages}{59--64}.
\bibitem[{Musaelian et~al.(2023)Musaelian, Batzner, Johansson, Sun, Owen, Kornbluth, and Kozinsky}]{MPNNs_2023}
\bibinfo{author}{A.~Musaelian}, \bibinfo{author}{S.~Batzner}, \bibinfo{author}{A.~Johansson}, \bibinfo{author}{L.~Sun}, \bibinfo{author}{C.~J. Owen}, \bibinfo{author}{M.~Kornbluth}, \bibinfo{author}{B.~Kozinsky},
\newblock \bibinfo{title}{Learning local equivariant representations for large-scale atomistic dynamics},
\newblock \bibinfo{journal}{Nature Communications} \bibinfo{volume}{14} (\bibinfo{year}{2023}) \bibinfo{pages}{579}.
\bibitem[{Guo et~al.(2022)Guo, Lu, Yan, Hu, Liu, Tan, Sun, Jiang, Liu, Chen, Zhang, Chen, Wang, and Jia}]{10.1145/3503221.3508425}
\bibinfo{author}{Z.~Guo}, \bibinfo{author}{D.~Lu}, \bibinfo{author}{Y.~Yan}, \bibinfo{author}{S.~Hu}, \bibinfo{author}{R.~Liu}, \bibinfo{author}{G.~Tan}, \bibinfo{author}{N.~Sun}, \bibinfo{author}{W.~Jiang}, \bibinfo{author}{L.~Liu}, \bibinfo{author}{Y.~Chen}, \bibinfo{author}{L.~Zhang}, \bibinfo{author}{M.~Chen}, \bibinfo{author}{H.~Wang}, \bibinfo{author}{W.~Jia},
\newblock \bibinfo{title}{Extending the limit of molecular dynamics with ab initio accuracy to 10 billion atoms},
\newblock in: \bibinfo{booktitle}{Proceedings of the 27th ACM SIGPLAN Symposium on Principles and Practice of Parallel Programming}, PPoPP '22, \bibinfo{publisher}{Association for Computing Machinery}, \bibinfo{address}{New York, NY, USA}, \bibinfo{year}{2022}, p. \bibinfo{pages}{205–218}.
\bibitem[{Doerr et~al.(2021)Doerr, Majewski, P{\'e}rez, Krämer, Clementi, Noe, Giorgino, and De~Fabritiis}]{doi:10.1021/acs.jctc.0c01343}
\bibinfo{author}{S.~Doerr}, \bibinfo{author}{M.~Majewski}, \bibinfo{author}{A.~P{\'e}rez}, \bibinfo{author}{A.~Krämer}, \bibinfo{author}{C.~Clementi}, \bibinfo{author}{F.~Noe}, \bibinfo{author}{T.~Giorgino}, \bibinfo{author}{G.~De~Fabritiis},
\newblock \bibinfo{title}{Torchmd: A deep learning framework for molecular simulations},
\newblock \bibinfo{journal}{Journal of Chemical Theory and Computation} \bibinfo{volume}{17} (\bibinfo{year}{2021}) \bibinfo{pages}{2355--2363}. \bibinfo{note}{PMID: 33729795}.
\bibitem[{Gao et~al.(2020)Gao, Ramezanghorbani, Isayev, Smith, and Roitberg}]{doi:10.1021/acs.jcim.0c00451}
\bibinfo{author}{X.~Gao}, \bibinfo{author}{F.~Ramezanghorbani}, \bibinfo{author}{O.~Isayev}, \bibinfo{author}{J.~S. Smith}, \bibinfo{author}{A.~E. Roitberg},
\newblock \bibinfo{title}{Torchani: A free and open source pytorch-based deep learning implementation of the ani neural network potentials},
\newblock \bibinfo{journal}{Journal of Chemical Information and Modeling} \bibinfo{volume}{60} (\bibinfo{year}{2020}) \bibinfo{pages}{3408--3415}. \bibinfo{note}{PMID: 32568524}.
\bibitem[{Kostiuchenko et~al.(2019)Kostiuchenko, K\"ormann, Neugebauer, and Shapeev}]{Korman_npj}
\bibinfo{author}{T.~Kostiuchenko}, \bibinfo{author}{F.~K\"ormann}, \bibinfo{author}{J.~Neugebauer}, \bibinfo{author}{A.~Shapeev},
\newblock \bibinfo{title}{Impact of lattice relaxations on phase transitions in a high-entropy alloy studied by machine-learning potentials},
\newblock \bibinfo{journal}{npj Computational Materials} \bibinfo{volume}{5} (\bibinfo{year}{2019}) \bibinfo{pages}{55}.
\bibitem[{Wang and Yang(2025)}]{WANG2025120635}
\bibinfo{author}{Z.~Wang}, \bibinfo{author}{T.~Yang},
\newblock \bibinfo{title}{Chemical order/disorder phase transitions in nicofealtib multi-principal element alloys: A monte carlo analysis},
\newblock \bibinfo{journal}{Acta Materialia} \bibinfo{volume}{285} (\bibinfo{year}{2025}) \bibinfo{pages}{120635}.
\bibitem[{Xie et~al.(2024)Xie, Zhou, Jin, and Jiang}]{doi:10.1021/acs.jctc.4c00463}
\bibinfo{author}{J.-Z. Xie}, \bibinfo{author}{X.-Y. Zhou}, \bibinfo{author}{B.~Jin}, \bibinfo{author}{H.~Jiang},
\newblock \bibinfo{title}{Machine learning force field-aided cluster expansion approach to phase diagram of alloyed materials},
\newblock \bibinfo{journal}{Journal of Chemical Theory and Computation} \bibinfo{volume}{20} (\bibinfo{year}{2024}) \bibinfo{pages}{6207--6217}. \bibinfo{note}{PMID: 38940547}.
\bibitem[{Jiang et~al.(2023)Jiang, Xie, and Wang}]{JIANG20231341}
\bibinfo{author}{D.~Jiang}, \bibinfo{author}{L.~Xie}, \bibinfo{author}{L.~Wang},
\newblock \bibinfo{title}{Current application status of multi-scale simulation and machine learning in research on high-entropy alloys},
\newblock \bibinfo{journal}{Journal of Materials Research and Technology} \bibinfo{volume}{26} (\bibinfo{year}{2023}) \bibinfo{pages}{1341--1374}.
\bibitem[{Körmann and Sluiter(2016)}]{e18080403}
\bibinfo{author}{F.~Körmann}, \bibinfo{author}{M.~H. Sluiter},
\newblock \bibinfo{title}{Interplay between lattice distortions, vibrations and phase stability in {NbMoTaW} high entropy alloys},
\newblock \bibinfo{journal}{Entropy} \bibinfo{volume}{18} (\bibinfo{year}{2016}).
\bibitem[{Liu et~al.(2021)Liu, Zhang, Yin, Bi, Eisenbach, and Wang}]{LIU2021110135}
\bibinfo{author}{X.~Liu}, \bibinfo{author}{J.~Zhang}, \bibinfo{author}{J.~Yin}, \bibinfo{author}{S.~Bi}, \bibinfo{author}{M.~Eisenbach}, \bibinfo{author}{Y.~Wang},
\newblock \bibinfo{title}{Monte carlo simulation of order-disorder transition in refractory high entropy alloys: A data-driven approach},
\newblock \bibinfo{journal}{Computational Materials Science} \bibinfo{volume}{187} (\bibinfo{year}{2021}) \bibinfo{pages}{110135}.
\bibitem[{Yin et~al.(2021)Yin, Pei, and Gao}]{Junqi-NCS}
\bibinfo{author}{J.~Yin}, \bibinfo{author}{Z.~Pei}, \bibinfo{author}{M.~C. Gao},
\newblock \bibinfo{title}{Neural network-based order parameter for phase transitions and its applications in high-entropy alloys},
\newblock \bibinfo{journal}{Nature Computational Science} \bibinfo{volume}{1} (\bibinfo{year}{2021}) \bibinfo{pages}{686--693}.
\bibitem[{Johnson(1986)}]{Johnson1986}
\bibinfo{author}{M.~A. Johnson}, \bibinfo{title}{Concurrent Computation and Its Application to the Study of Melting in Two Dimensions}, \bibinfo{type}{Thesis}, California Institute of Technology, \bibinfo{year}{1986}.
\bibitem[{Heffelfinger and Lewitt(1996)}]{1996-JCC}
\bibinfo{author}{G.~S. Heffelfinger}, \bibinfo{author}{M.~E. Lewitt},
\newblock \bibinfo{title}{A comparison between two massively parallel algorithms for monte carlo computer simulation: An investigation in the grand canonical ensemble},
\newblock \bibinfo{journal}{Journal of Computational Chemistry} \bibinfo{volume}{17} (\bibinfo{year}{1996}) \bibinfo{pages}{250--265}.
\bibitem[{Preis et~al.(2009)Preis, Virnau, Paul, and Schneider}]{PREIS20094468}
\bibinfo{author}{T.~Preis}, \bibinfo{author}{P.~Virnau}, \bibinfo{author}{W.~Paul}, \bibinfo{author}{J.~J. Schneider},
\newblock \bibinfo{title}{Gpu accelerated monte carlo simulation of the 2d and 3d ising model},
\newblock \bibinfo{journal}{Journal of Computational Physics} \bibinfo{volume}{228} (\bibinfo{year}{2009}) \bibinfo{pages}{4468--4477}.
\bibitem[{Block et~al.(2010)Block, Virnau, and Preis}]{BLOCK20101549}
\bibinfo{author}{B.~Block}, \bibinfo{author}{P.~Virnau}, \bibinfo{author}{T.~Preis},
\newblock \bibinfo{title}{Multi-gpu accelerated multi-spin monte carlo simulations of the 2d ising model},
\newblock \bibinfo{journal}{Computer Physics Communications} \bibinfo{volume}{181} (\bibinfo{year}{2010}) \bibinfo{pages}{1549--1556}.
\bibitem[{Sadigh et~al.(2012)Sadigh, Erhart, Stukowski, Caro, Martinez, and Zepeda-Ruiz}]{sadigh2012scalable}
\bibinfo{author}{B.~Sadigh}, \bibinfo{author}{P.~Erhart}, \bibinfo{author}{A.~Stukowski}, \bibinfo{author}{A.~Caro}, \bibinfo{author}{E.~Martinez}, \bibinfo{author}{L.~Zepeda-Ruiz},
\newblock \bibinfo{title}{Scalable parallel monte carlo algorithm for atomistic simulations of precipitation in alloys},
\newblock \bibinfo{journal}{Physical Review B} \bibinfo{volume}{85} (\bibinfo{year}{2012}) \bibinfo{pages}{184203}.
\bibitem[{Mick et~al.(2013)Mick, Hailat, Russo, Rushaidat, Schwiebert, and Potoff}]{mick2013gpu}
\bibinfo{author}{J.~Mick}, \bibinfo{author}{E.~Hailat}, \bibinfo{author}{V.~Russo}, \bibinfo{author}{K.~Rushaidat}, \bibinfo{author}{L.~Schwiebert}, \bibinfo{author}{J.~Potoff},
\newblock \bibinfo{title}{Gpu‑accelerated gibbs ensemble monte carlo simulations of lennard‑jonesium},
\newblock \bibinfo{journal}{Computer Physics Communications} \bibinfo{volume}{184} (\bibinfo{year}{2013}) \bibinfo{pages}{2662--2669}.
\bibitem[{Ortega-Zamorano et~al.(2013)Ortega-Zamorano, Montemurro, Cannas, Jerez, and Franco}]{OrtegaZamorano2013FPGAIsing}
\bibinfo{author}{F.~Ortega-Zamorano}, \bibinfo{author}{M.~A. Montemurro}, \bibinfo{author}{S.~A. Cannas}, \bibinfo{author}{J.~M. Jerez}, \bibinfo{author}{L.~Franco},
\newblock \bibinfo{title}{Monte carlo simulation of the ising model on fpga},
\newblock \bibinfo{journal}{Journal of Computational Physics} \bibinfo{volume}{237} (\bibinfo{year}{2013}) \bibinfo{pages}{224--234}.
\bibitem[{Anderson et~al.(2013)Anderson, Jankowski, Grubb, Engel, and Glotzer}]{anderson2013massively}
\bibinfo{author}{J.~A. Anderson}, \bibinfo{author}{E.~Jankowski}, \bibinfo{author}{T.~L. Grubb}, \bibinfo{author}{M.~Engel}, \bibinfo{author}{S.~C. Glotzer},
\newblock \bibinfo{title}{Massively parallel monte carlo for many-particle simulations on gpus},
\newblock \bibinfo{journal}{Journal of Computational Physics} \bibinfo{volume}{254} (\bibinfo{year}{2013}) \bibinfo{pages}{27--38}.
\bibitem[{Anderson and Glotzer(2016)}]{Anderson2016SMC}
\bibinfo{author}{J.~A. Anderson}, \bibinfo{author}{S.~C. Glotzer},
\newblock \bibinfo{title}{Scalable metropolis monte carlo for simulation of hard shapes},
\newblock \bibinfo{journal}{Computer Physics Communications} \bibinfo{volume}{204} (\bibinfo{year}{2016}) \bibinfo{pages}{21--30}.
\bibitem[{Liang et~al.(2017)Liang, Xing, and Li}]{Liang2017GpuMonteCarlo}
\bibinfo{author}{Y.~Liang}, \bibinfo{author}{X.~Xing}, \bibinfo{author}{Y.~Li},
\newblock \bibinfo{title}{A gpu-based large-scale monte carlo simulation method for systems with long-range interactions},
\newblock \bibinfo{journal}{Journal of Computational Physics} \bibinfo{volume}{338} (\bibinfo{year}{2017}) \bibinfo{pages}{252--268}.
\bibitem[{Liu et~al.(2019)Liu, Wang, Wang, Wu, Li, Zhang, Guo, Lin, Xie, Xiong et~al.}]{liu2019high}
\bibinfo{author}{L.~Liu}, \bibinfo{author}{Z.~Wang}, \bibinfo{author}{Z.~Wang}, \bibinfo{author}{J.~Wu}, \bibinfo{author}{M.~Li}, \bibinfo{author}{Z.~Zhang}, \bibinfo{author}{Q.~Guo}, \bibinfo{author}{X.~Lin}, \bibinfo{author}{Y.~Xie}, \bibinfo{author}{X.~Xiong}, et~al.,
\newblock \bibinfo{title}{High performance monte carlo simulation of ising model on tpu clusters},
\newblock in: \bibinfo{booktitle}{Proceedings of the International Conference for High Performance Computing, Networking, Storage and Analysis}, \bibinfo{organization}{Association for Computing Machinery}, \bibinfo{publisher}{ACM}, \bibinfo{address}{New York, NY, USA}, \bibinfo{year}{2019}, pp. \bibinfo{pages}{1--20}.
\bibitem[{Romero et~al.(2020)Romero, Bisson, Fatica, and Bernaschi}]{ROMERO2020107473}
\bibinfo{author}{J.~Romero}, \bibinfo{author}{M.~Bisson}, \bibinfo{author}{M.~Fatica}, \bibinfo{author}{M.~Bernaschi},
\newblock \bibinfo{title}{High performance implementations of the 2d ising model on gpus},
\newblock \bibinfo{journal}{Computer Physics Communications} \bibinfo{volume}{256} (\bibinfo{year}{2020}) \bibinfo{pages}{107473}.
\bibitem[{Lepadatu(2023)}]{lepadatu2023accelerating}
\bibinfo{author}{S.~Lepadatu},
\newblock \bibinfo{title}{Accelerating micromagnetic and atomistic simulations using multiple gpus},
\newblock \bibinfo{journal}{Journal of Applied Physics} \bibinfo{volume}{134} (\bibinfo{year}{2023}) \bibinfo{pages}{163903}.
\bibitem[{Thompson et~al.(2022)Thompson, Aktulga, Berger, Bolintineanu, Brown, Crozier, {in 't Veld}, Kohlmeyer, Moore, Nguyen, Shan, Stevens, Tranchida, Trott, and Plimpton}]{THOMPSON2022108171}
\bibinfo{author}{A.~P. Thompson}, \bibinfo{author}{H.~M. Aktulga}, \bibinfo{author}{R.~Berger}, \bibinfo{author}{D.~S. Bolintineanu}, \bibinfo{author}{W.~M. Brown}, \bibinfo{author}{P.~S. Crozier}, \bibinfo{author}{P.~J. {in 't Veld}}, \bibinfo{author}{A.~Kohlmeyer}, \bibinfo{author}{S.~G. Moore}, \bibinfo{author}{T.~D. Nguyen}, \bibinfo{author}{R.~Shan}, \bibinfo{author}{M.~J. Stevens}, \bibinfo{author}{J.~Tranchida}, \bibinfo{author}{C.~Trott}, \bibinfo{author}{S.~J. Plimpton},
\newblock \bibinfo{title}{Lammps - a flexible simulation tool for particle-based materials modeling at the atomic, meso, and continuum scales},
\newblock \bibinfo{journal}{Computer Physics Communications} \bibinfo{volume}{271} (\bibinfo{year}{2022}) \bibinfo{pages}{108171}.
\bibitem[{Li et~al.(2020)Li, Chen, Zheng, Zuo, and Ong}]{2020NPJ_Li}
\bibinfo{author}{X.-G. Li}, \bibinfo{author}{C.~Chen}, \bibinfo{author}{H.~Zheng}, \bibinfo{author}{Y.~Zuo}, \bibinfo{author}{S.~P. Ong},
\newblock \bibinfo{title}{{Complex strengthening mechanisms in the NbMoTaW multi-principal element alloy}},
\newblock \bibinfo{journal}{npj Computational Materials} \bibinfo{volume}{6} (\bibinfo{year}{2020}) \bibinfo{pages}{70}.
\bibitem[{Yin et~al.(2021)Yin, Zuo, Abu-Odeh, Zheng, Li, Ding, Ong, Asta, and Ritchie}]{Yin2021}
\bibinfo{author}{S.~Yin}, \bibinfo{author}{Y.~Zuo}, \bibinfo{author}{A.~Abu-Odeh}, \bibinfo{author}{H.~Zheng}, \bibinfo{author}{X.-G. Li}, \bibinfo{author}{J.~Ding}, \bibinfo{author}{S.~P. Ong}, \bibinfo{author}{M.~Asta}, \bibinfo{author}{R.~O. Ritchie},
\newblock \bibinfo{title}{Atomistic simulations of dislocation mobility in refractory high-entropy alloys and the effect of chemical short-range order},
\newblock \bibinfo{journal}{Nature Communications} \bibinfo{volume}{12} (\bibinfo{year}{2021}) \bibinfo{pages}{4873}.
\bibitem[{Yeh et~al.(2004)Yeh, Chen, Lin, Gan, Chin, Shun, Tsau, and Chang}]{ADEM:ADEM200300567}
\bibinfo{author}{J.-W. Yeh}, \bibinfo{author}{S.-K. Chen}, \bibinfo{author}{S.-J. Lin}, \bibinfo{author}{J.-Y. Gan}, \bibinfo{author}{T.-S. Chin}, \bibinfo{author}{T.-T. Shun}, \bibinfo{author}{C.-H. Tsau}, \bibinfo{author}{S.-Y. Chang},
\newblock \bibinfo{title}{Nanostructured high-entropy alloys with multiple principal elements: Novel alloy design concepts and outcomes},
\newblock \bibinfo{journal}{Advanced Engineering Materials} \bibinfo{volume}{6} (\bibinfo{year}{2004}) \bibinfo{pages}{299--303}.
\bibitem[{Cantor et~al.(2004)Cantor, Chang, Knight, and Vincent}]{CANTOR2004213}
\bibinfo{author}{B.~Cantor}, \bibinfo{author}{I.~Chang}, \bibinfo{author}{P.~Knight}, \bibinfo{author}{A.~Vincent},
\newblock \bibinfo{title}{Microstructural development in equiatomic multicomponent alloys},
\newblock \bibinfo{journal}{Materials Science and Engineering: A} \bibinfo{volume}{375-377} (\bibinfo{year}{2004}) \bibinfo{pages}{213 -- 218}.
\bibitem[{Li et~al.(2016)Li, Pradeep, Deng, Raabe, and Tasan}]{NatureRaabe}
\bibinfo{author}{Z.~Li}, \bibinfo{author}{K.~G. Pradeep}, \bibinfo{author}{Y.~Deng}, \bibinfo{author}{D.~Raabe}, \bibinfo{author}{C.~C. Tasan},
\newblock \bibinfo{title}{{Metastable high-entropy dual-phase alloys overcome the strength–ductility trade-off}},
\newblock \bibinfo{journal}{Nature} \bibinfo{volume}{534} (\bibinfo{year}{2016}) \bibinfo{pages}{227}.
\bibitem[{Yang et~al.(2018)Yang, Zhao, Tong, Jiao, Wei, Cai, Han, Chen, Hu, Kai, Lu, Liu, and Liu}]{Yang933}
\bibinfo{author}{T.~Yang}, \bibinfo{author}{Y.~L. Zhao}, \bibinfo{author}{Y.~Tong}, \bibinfo{author}{Z.~B. Jiao}, \bibinfo{author}{J.~Wei}, \bibinfo{author}{J.~X. Cai}, \bibinfo{author}{X.~D. Han}, \bibinfo{author}{D.~Chen}, \bibinfo{author}{A.~Hu}, \bibinfo{author}{J.~J. Kai}, \bibinfo{author}{K.~Lu}, \bibinfo{author}{Y.~Liu}, \bibinfo{author}{C.~T. Liu},
\newblock \bibinfo{title}{Multicomponent intermetallic nanoparticles and superb mechanical behaviors of complex alloys},
\newblock \bibinfo{journal}{Science} \bibinfo{volume}{362} (\bibinfo{year}{2018}) \bibinfo{pages}{933--937}.
\bibitem[{Yang et~al.(2021)Yang, Chen, Tan, Poplawsky, An, Wang, Samolyuk, Littrell, Lupini, Borisevich, and George}]{ORNL_HEAs_2021}
\bibinfo{author}{Y.~Yang}, \bibinfo{author}{T.~Chen}, \bibinfo{author}{L.~Tan}, \bibinfo{author}{J.~D. Poplawsky}, \bibinfo{author}{K.~An}, \bibinfo{author}{Y.~Wang}, \bibinfo{author}{G.~D. Samolyuk}, \bibinfo{author}{K.~Littrell}, \bibinfo{author}{A.~R. Lupini}, \bibinfo{author}{A.~Borisevich}, \bibinfo{author}{E.~P. George},
\newblock \bibinfo{title}{Bifunctional nanoprecipitates strengthen and ductilize a medium-entropy alloy},
\newblock \bibinfo{journal}{Nature} \bibinfo{volume}{595} (\bibinfo{year}{2021}) \bibinfo{pages}{245--249}.
\bibitem[{Ferrari et~al.(2023)Ferrari, K{\"o}rmann, Asta, and Neugebauer}]{SRO_NCS_2023}
\bibinfo{author}{A.~Ferrari}, \bibinfo{author}{F.~K{\"o}rmann}, \bibinfo{author}{M.~Asta}, \bibinfo{author}{J.~Neugebauer},
\newblock \bibinfo{title}{Simulating short-range order in compositionally complex materials},
\newblock \bibinfo{journal}{Nature Computational Science} \bibinfo{volume}{3} (\bibinfo{year}{2023}) \bibinfo{pages}{221--229}.
\bibitem[{He et~al.(2024)He, Davids, Breen, and Ringer}]{APT_CoCrNi_NatureMat_2024}
\bibinfo{author}{M.~He}, \bibinfo{author}{W.~J. Davids}, \bibinfo{author}{A.~J. Breen}, \bibinfo{author}{S.~P. Ringer},
\newblock \bibinfo{title}{Quantifying short-range order using atom probe tomography},
\newblock \bibinfo{journal}{Nature Materials} \bibinfo{volume}{23} (\bibinfo{year}{2024}) \bibinfo{pages}{1200--1207}.
\bibitem[{Xiao et~al.(2022)Xiao, Luan, Zhao, Zhang, Chen, Zhao, Xu, Liu, Kai, and Yang}]{NanoParticleHEA2022}
\bibinfo{author}{B.~Xiao}, \bibinfo{author}{J.~Luan}, \bibinfo{author}{S.~Zhao}, \bibinfo{author}{L.~Zhang}, \bibinfo{author}{S.~Chen}, \bibinfo{author}{Y.~Zhao}, \bibinfo{author}{L.~Xu}, \bibinfo{author}{C.~T. Liu}, \bibinfo{author}{J.-J. Kai}, \bibinfo{author}{T.~Yang},
\newblock \bibinfo{title}{Achieving thermally stable nanoparticles in chemically complex alloys via controllable sluggish lattice diffusion},
\newblock \bibinfo{journal}{Nature Communications} \bibinfo{volume}{13} (\bibinfo{year}{2022}) \bibinfo{pages}{4870}.
\bibitem[{Han et~al.(2023)Han, Wu, Zhao, Guo, Yan, Hong, and Wang}]{HAN20231717}
\bibinfo{author}{X.~Han}, \bibinfo{author}{G.~Wu}, \bibinfo{author}{S.~Zhao}, \bibinfo{author}{J.~Guo}, \bibinfo{author}{M.~Yan}, \bibinfo{author}{X.~Hong}, \bibinfo{author}{D.~Wang},
\newblock \bibinfo{title}{Nanoscale high-entropy alloy for electrocatalysis},
\newblock \bibinfo{journal}{Matter} \bibinfo{volume}{6} (\bibinfo{year}{2023}) \bibinfo{pages}{1717--1751}.
\bibitem[{Li et~al.(2024)Li, Lin, Zhang, Zhao, Tao, Li, Li, Zeng, Luo, and Guo}]{doi:10.1126/sciadv.adn2877}
\bibinfo{author}{M.~Li}, \bibinfo{author}{F.~Lin}, \bibinfo{author}{S.~Zhang}, \bibinfo{author}{R.~Zhao}, \bibinfo{author}{L.~Tao}, \bibinfo{author}{L.~Li}, \bibinfo{author}{J.~Li}, \bibinfo{author}{L.~Zeng}, \bibinfo{author}{M.~Luo}, \bibinfo{author}{S.~Guo},
\newblock \bibinfo{title}{High-entropy alloy electrocatalysts go to (sub-)nanoscale},
\newblock \bibinfo{journal}{Science Advances} \bibinfo{volume}{10} (\bibinfo{year}{2024}) \bibinfo{pages}{eadn2877}.
\bibitem[{Yao et~al.(2022)Yao, Dong, Brozena, Luo, Miao, Chi, Wang, Kevrekidis, Ren, Greeley, Wang, Anapolsky, and Hu}]{doi:10.1126/science.abn3103}
\bibinfo{author}{Y.~Yao}, \bibinfo{author}{Q.~Dong}, \bibinfo{author}{A.~Brozena}, \bibinfo{author}{J.~Luo}, \bibinfo{author}{J.~Miao}, \bibinfo{author}{M.~Chi}, \bibinfo{author}{C.~Wang}, \bibinfo{author}{I.~G. Kevrekidis}, \bibinfo{author}{Z.~J. Ren}, \bibinfo{author}{J.~Greeley}, \bibinfo{author}{G.~Wang}, \bibinfo{author}{A.~Anapolsky}, \bibinfo{author}{L.~Hu},
\newblock \bibinfo{title}{High-entropy nanoparticles: Synthesis-structure-property relationships and data-driven discovery},
\newblock \bibinfo{journal}{Science} \bibinfo{volume}{376} (\bibinfo{year}{2022}) \bibinfo{pages}{eabn3103}.
\bibitem[{Zhang et~al.(2020)Zhang, Liu, Bi, Yin, Zhang, and Eisenbach}]{ZHANG2020108247}
\bibinfo{author}{J.~Zhang}, \bibinfo{author}{X.~Liu}, \bibinfo{author}{S.~Bi}, \bibinfo{author}{J.~Yin}, \bibinfo{author}{G.~Zhang}, \bibinfo{author}{M.~Eisenbach},
\newblock \bibinfo{title}{Robust data-driven approach for predicting the configurational energy of high entropy alloys},
\newblock \bibinfo{journal}{Materials \& Design} \bibinfo{volume}{185} (\bibinfo{year}{2020}) \bibinfo{pages}{108247}.
\bibitem[{Liu et~al.(2024)Liu, Wang, Luan, Chen, Cai, Chen, Lu, Fan, Yu, and Chou}]{doi:10.1021/acs.jctc.4c00340}
\bibinfo{author}{J.~Liu}, \bibinfo{author}{P.~Wang}, \bibinfo{author}{J.~Luan}, \bibinfo{author}{J.~Chen}, \bibinfo{author}{P.~Cai}, \bibinfo{author}{J.~Chen}, \bibinfo{author}{X.~Lu}, \bibinfo{author}{Y.~Fan}, \bibinfo{author}{Z.~Yu}, \bibinfo{author}{K.~Chou},
\newblock \bibinfo{title}{Vase: A high-entropy alloy short-range order structural descriptor for machine learning},
\newblock \bibinfo{journal}{Journal of Chemical Theory and Computation} \bibinfo{volume}{20} (\bibinfo{year}{2024}) \bibinfo{pages}{11082--11092}. \bibinfo{note}{PMID: 39046791}.
\bibitem[{Wang et~al.(1995)Wang, Stocks, Shelton, Nicholson, Szotek, and Temmerman}]{PhysRevLett.75.2867}
\bibinfo{author}{Y.~Wang}, \bibinfo{author}{G.~M. Stocks}, \bibinfo{author}{W.~A. Shelton}, \bibinfo{author}{D.~M.~C. Nicholson}, \bibinfo{author}{Z.~Szotek}, \bibinfo{author}{W.~M. Temmerman},
\newblock \bibinfo{title}{Order-{N} multiple scattering approach to electronic structure calculations},
\newblock \bibinfo{journal}{Phys. Rev. Lett.} \bibinfo{volume}{75} (\bibinfo{year}{1995}) \bibinfo{pages}{2867--2870}.
\bibitem[{Eisenbach et~al.(2017)Eisenbach, Li, Liu, Odbadrakh, Pei, Stocks, and Yin}]{osti_1420087}
\bibinfo{author}{M.~Eisenbach}, \bibinfo{author}{Y.~W. Li}, \bibinfo{author}{X.~Liu}, \bibinfo{author}{O.~K. Odbadrakh}, \bibinfo{author}{Z.~Pei}, \bibinfo{author}{G.~M. Stocks}, \bibinfo{author}{J.~Yin}, \bibinfo{title}{Lsms}, \bibinfo{year}{2017}.
\bibitem[{Li et~al.(2016)Li, Pradeep, Deng, Raabe, and Tasan}]{NanophaseHEA}
\bibinfo{author}{Z.~Li}, \bibinfo{author}{K.~G. Pradeep}, \bibinfo{author}{Y.~Deng}, \bibinfo{author}{D.~Raabe}, \bibinfo{author}{C.~C. Tasan},
\newblock \bibinfo{title}{Metastable high-entropy dual-phase alloys overcome the strength--ductility trade-off},
\newblock \bibinfo{journal}{Nature} \bibinfo{volume}{534} (\bibinfo{year}{2016}) \bibinfo{pages}{227--230}.
\bibitem[{Fu et~al.(2018)Fu, Jiang, Wardini, MacDonald, Wen, Xiong, Zhang, Zhou, Rupert, Chen, and Lavernia}]{Fueaat8712}
\bibinfo{author}{Z.~Fu}, \bibinfo{author}{L.~Jiang}, \bibinfo{author}{J.~L. Wardini}, \bibinfo{author}{B.~E. MacDonald}, \bibinfo{author}{H.~Wen}, \bibinfo{author}{W.~Xiong}, \bibinfo{author}{D.~Zhang}, \bibinfo{author}{Y.~Zhou}, \bibinfo{author}{T.~J. Rupert}, \bibinfo{author}{W.~Chen}, \bibinfo{author}{E.~J. Lavernia},
\newblock \bibinfo{title}{A high-entropy alloy with hierarchical nanoprecipitates and ultrahigh strength},
\newblock \bibinfo{journal}{Science Advances} \bibinfo{volume}{4} (\bibinfo{year}{2018}).
\bibitem[{Santodonato et~al.(2018)Santodonato, Liaw, Unocic, Bei, and Morris}]{Santodonato2018}
\bibinfo{author}{L.~J. Santodonato}, \bibinfo{author}{P.~K. Liaw}, \bibinfo{author}{R.~R. Unocic}, \bibinfo{author}{H.~Bei}, \bibinfo{author}{J.~R. Morris},
\newblock \bibinfo{title}{Predictive multiphase evolution in {Al}-containing high-entropy alloys},
\newblock \bibinfo{journal}{Nature Communications} \bibinfo{volume}{9} (\bibinfo{year}{2018}) \bibinfo{pages}{4520}.
\bibitem[{Yao et~al.(2024)Yao, Cappola, Zhang, Zhu, Cai, Yu, and Li}]{YAO2024120457}
\bibinfo{author}{Y.~Yao}, \bibinfo{author}{J.~Cappola}, \bibinfo{author}{Z.~Zhang}, \bibinfo{author}{Q.~Zhu}, \bibinfo{author}{W.~Cai}, \bibinfo{author}{X.~Yu}, \bibinfo{author}{L.~Li},
\newblock \bibinfo{title}{Nanostructure and dislocation interactions in refractory complex concentrated alloy: From chemical short-range order to nanoscale b2 precipitates},
\newblock \bibinfo{journal}{Acta Materialia} \bibinfo{volume}{281} (\bibinfo{year}{2024}) \bibinfo{pages}{120457}.
\bibitem[{Wang et~al.(2025)Wang, Meng, Guo, Li, Liu, Li, Xiao, Zhao, Sun, Tan, and Jia}]{10880101}
\bibinfo{author}{X.~Wang}, \bibinfo{author}{X.~Meng}, \bibinfo{author}{Z.~Guo}, \bibinfo{author}{M.~Li}, \bibinfo{author}{L.~Liu}, \bibinfo{author}{M.~Li}, \bibinfo{author}{Q.~Xiao}, \bibinfo{author}{T.~Zhao}, \bibinfo{author}{N.~Sun}, \bibinfo{author}{G.~Tan}, \bibinfo{author}{W.~Jia},
\newblock \bibinfo{title}{29-billion atoms molecular dynamics simulation with ab initio accuracy on 35 million cores of new sunway supercomputer},
\newblock \bibinfo{journal}{IEEE Transactions on Computers}  (\bibinfo{year}{2025}) \bibinfo{pages}{1--14}.
\bibitem[{Liu et~al.(2019)Liu, Zhang, Eisenbach, and Wang}]{liu2019machine}
\bibinfo{author}{X.~Liu}, \bibinfo{author}{J.~Zhang}, \bibinfo{author}{M.~Eisenbach}, \bibinfo{author}{Y.~Wang},
\newblock \bibinfo{title}{Machine learning modeling of high entropy alloy: the role of short-range order},
\newblock \bibinfo{journal}{arXiv preprint arXiv:1906.02889}  (\bibinfo{year}{2019}).
\bibitem[{Kirklin et~al.(2015)Kirklin, Saal, Meredig, Thompson, Doak, Aykol, R{\"u}hl, and Wolverton}]{OQMD}
\bibinfo{author}{S.~Kirklin}, \bibinfo{author}{J.~E. Saal}, \bibinfo{author}{B.~Meredig}, \bibinfo{author}{A.~Thompson}, \bibinfo{author}{J.~W. Doak}, \bibinfo{author}{M.~Aykol}, \bibinfo{author}{S.~R{\"u}hl}, \bibinfo{author}{C.~Wolverton},
\newblock \bibinfo{title}{The open quantum materials database (oqmd): assessing the accuracy of dft formation energies},
\newblock \bibinfo{journal}{npj Computational Materials} \bibinfo{volume}{1} (\bibinfo{year}{2015}) \bibinfo{pages}{15010}.
\bibitem[{Huhn and Widom(2013)}]{Huhn2013}
\bibinfo{author}{W.~P. Huhn}, \bibinfo{author}{M.~Widom},
\newblock \bibinfo{title}{Prediction of {A2 to B2} phase transition in the high-entropy alloy {Mo-Nb-Ta-W}},
\newblock \bibinfo{journal}{JOM} \bibinfo{volume}{65} (\bibinfo{year}{2013}) \bibinfo{pages}{1772--1779}.
\bibitem[{Zhang et~al.(2022)Zhang, Tang, Wen, Obaied, Roslyakova, and Zhang}]{ZHANG2022105780}
\bibinfo{author}{E.~Zhang}, \bibinfo{author}{Y.~Tang}, \bibinfo{author}{M.~Wen}, \bibinfo{author}{A.~Obaied}, \bibinfo{author}{I.~Roslyakova}, \bibinfo{author}{L.~Zhang},
\newblock \bibinfo{title}{On phase stability of mo-nb-ta-w refractory high entropy alloys},
\newblock \bibinfo{journal}{International Journal of Refractory Metals and Hard Materials} \bibinfo{volume}{103} (\bibinfo{year}{2022}) \bibinfo{pages}{105780}.
\bibitem[{Fu et~al.(2019)Fu, Hoffman, MacDonald, Jiang, Chen, Arivu, Wen, and Lavernia}]{FU2019372}
\bibinfo{author}{Z.~Fu}, \bibinfo{author}{A.~Hoffman}, \bibinfo{author}{B.~E. MacDonald}, \bibinfo{author}{Z.~Jiang}, \bibinfo{author}{W.~Chen}, \bibinfo{author}{M.~Arivu}, \bibinfo{author}{H.~Wen}, \bibinfo{author}{E.~J. Lavernia},
\newblock \bibinfo{title}{Atom probe tomography study of an fe25ni25co25ti15al10 high-entropy alloy fabricated by powder metallurgy},
\newblock \bibinfo{journal}{Acta Materialia} \bibinfo{volume}{179} (\bibinfo{year}{2019}) \bibinfo{pages}{372--382}.
\bibitem[{Liao et~al.(2025)Liao, Kuroki, Tamaki, Arao, Matsumoto, Imai, and Yamaguchi}]{3DCNP-ACS}
\bibinfo{author}{Q.~Liao}, \bibinfo{author}{H.~Kuroki}, \bibinfo{author}{T.~Tamaki}, \bibinfo{author}{M.~Arao}, \bibinfo{author}{M.~Matsumoto}, \bibinfo{author}{H.~Imai}, \bibinfo{author}{T.~Yamaguchi},
\newblock \bibinfo{title}{Three-dimensionally connected platinum--cobalt nanoparticles as support-free electrocatalysts for oxygen reduction},
\newblock \bibinfo{journal}{ACS Applied Nano Materials} \bibinfo{volume}{8} (\bibinfo{year}{2025}) \bibinfo{pages}{3323--3332}.
\bibitem[{Zhu et~al.(2024)Zhu, Gao, Yao, Hu, Li, Teng, Wang, Gong, Chen, and Yang}]{HEA_Nano_ACS_Nano}
\bibinfo{author}{W.~Zhu}, \bibinfo{author}{X.~Gao}, \bibinfo{author}{Y.~Yao}, \bibinfo{author}{S.~Hu}, \bibinfo{author}{Z.~Li}, \bibinfo{author}{Y.~Teng}, \bibinfo{author}{H.~Wang}, \bibinfo{author}{H.~Gong}, \bibinfo{author}{Z.~Chen}, \bibinfo{author}{Y.~Yang},
\newblock \bibinfo{title}{Nanostructured high entropy alloys as structural and functional materials},
\newblock \bibinfo{journal}{ACS Nano} \bibinfo{volume}{18} (\bibinfo{year}{2024}) \bibinfo{pages}{12672--12706}.
\bibitem[{Körmann et~al.(2017)Körmann, Ruban, and and}]{Körmann02012017}
\bibinfo{author}{F.~Körmann}, \bibinfo{author}{A.~V. Ruban}, \bibinfo{author}{M.~H.~S. and},
\newblock \bibinfo{title}{Long-ranged interactions in bcc nbmotaw high-entropy alloys},
\newblock \bibinfo{journal}{Materials Research Letters} \bibinfo{volume}{5} (\bibinfo{year}{2017}) \bibinfo{pages}{35--40}.
\bibitem[{Laplanche et~al.(2018)Laplanche, Gadaud, Bärsch, Demtröder, Reinhart, Schreuer, and George}]{LAPLANCHE2018244}
\bibinfo{author}{G.~Laplanche}, \bibinfo{author}{P.~Gadaud}, \bibinfo{author}{C.~Bärsch}, \bibinfo{author}{K.~Demtröder}, \bibinfo{author}{C.~Reinhart}, \bibinfo{author}{J.~Schreuer}, \bibinfo{author}{E.~George},
\newblock \bibinfo{title}{Elastic moduli and thermal expansion coefficients of medium-entropy subsystems of the crmnfeconi high-entropy alloy},
\newblock \bibinfo{journal}{Journal of Alloys and Compounds} \bibinfo{volume}{746} (\bibinfo{year}{2018}) \bibinfo{pages}{244--255}.
\bibitem[{Liu et~al.(2018)Liu, Wang, Eisenbach, and Stocks}]{LIU2018265}
\bibinfo{author}{X.~Liu}, \bibinfo{author}{Y.~Wang}, \bibinfo{author}{M.~Eisenbach}, \bibinfo{author}{G.~M. Stocks},
\newblock \bibinfo{title}{Fully-relativistic full-potential multiple scattering theory: A pathology-free scheme},
\newblock \bibinfo{journal}{Computer Physics Communications} \bibinfo{volume}{224} (\bibinfo{year}{2018}) \bibinfo{pages}{265--272}.
\bibitem[{Prodan and Kohn(2005)}]{doi:10.1073/pnas.0505436102}
\bibinfo{author}{E.~Prodan}, \bibinfo{author}{W.~Kohn},
\newblock \bibinfo{title}{Nearsightedness of electronic matter},
\newblock \bibinfo{journal}{Proceedings of the National Academy of Sciences} \bibinfo{volume}{102} (\bibinfo{year}{2005}) \bibinfo{pages}{11635--11638}.
\bibitem[{Lulli et~al.(2015)Lulli, Bernaschi, and Parisi}]{LULLI2015290}
\bibinfo{author}{M.~Lulli}, \bibinfo{author}{M.~Bernaschi}, \bibinfo{author}{G.~Parisi},
\newblock \bibinfo{title}{Highly optimized simulations on single- and multi-gpu systems of the 3d ising spin glass model},
\newblock \bibinfo{journal}{Computer Physics Communications} \bibinfo{volume}{196} (\bibinfo{year}{2015}) \bibinfo{pages}{290--303}.
\bibitem[{Yang et~al.(2019)Yang, Chen, Roumpos, Colby, and Anderson}]{10.1145/3295500.3356149}
\bibinfo{author}{K.~Yang}, \bibinfo{author}{Y.-F. Chen}, \bibinfo{author}{G.~Roumpos}, \bibinfo{author}{C.~Colby}, \bibinfo{author}{J.~Anderson},
\newblock \bibinfo{title}{High performance monte carlo simulation of ising model on tpu clusters},
\newblock in: \bibinfo{booktitle}{Proceedings of the International Conference for High Performance Computing, Networking, Storage and Analysis}, SC '19, \bibinfo{publisher}{Association for Computing Machinery}, \bibinfo{address}{New York, NY, USA}, \bibinfo{year}{2019}.
\bibitem[{Lin et~al.(2013)Lin, Wang, Zheng, Gao, and Zhang}]{LIN2013224}
\bibinfo{author}{Y.~Lin}, \bibinfo{author}{F.~Wang}, \bibinfo{author}{X.~Zheng}, \bibinfo{author}{H.~Gao}, \bibinfo{author}{L.~Zhang},
\newblock \bibinfo{title}{Monte carlo simulation of the ising model on fpga},
\newblock \bibinfo{journal}{Journal of Computational Physics} \bibinfo{volume}{237} (\bibinfo{year}{2013}) \bibinfo{pages}{224--234}.
\bibitem[{Ortega-Zamorano et~al.(2016)Ortega-Zamorano, Montemurro, Cannas, Jerez, and Franco}]{10.1109/TPDS.2015.2505725}
\bibinfo{author}{F.~Ortega-Zamorano}, \bibinfo{author}{M.~A. Montemurro}, \bibinfo{author}{S.~A. Cannas}, \bibinfo{author}{J.~M. Jerez}, \bibinfo{author}{L.~Franco},
\newblock \bibinfo{title}{Fpga hardware acceleration of monte carlo simulations for the ising model},
\newblock \bibinfo{journal}{IEEE Trans. Parallel Distrib. Syst.} \bibinfo{volume}{27} (\bibinfo{year}{2016}) \bibinfo{pages}{2618–2627}.
\bibitem[{Yamakov(2016)}]{Yamakov2016ParallelGC}
\bibinfo{author}{V.~Yamakov},
\newblock \bibinfo{title}{Parallel grand canonical monte carlo (paragrandmc) simulation code}.
\bibitem[{Momma and Izumi(2011)}]{VESTA}
\bibinfo{author}{K.~Momma}, \bibinfo{author}{F.~Izumi},
\newblock \bibinfo{title}{Vesta 3 for three-dimensional visualization of crystal, volumetric and morphology data},
\newblock \bibinfo{journal}{Journal of Applied Crystallography} \bibinfo{volume}{44} (\bibinfo{year}{2011}) \bibinfo{pages}{1272--1276}.
\bibitem[{Onsager(1944)}]{PhysRev.65.117}
\bibinfo{author}{L.~Onsager},
\newblock \bibinfo{title}{Crystal statistics. i. a two-dimensional model with an order-disorder transition},
\newblock \bibinfo{journal}{Phys. Rev.} \bibinfo{volume}{65} (\bibinfo{year}{1944}) \bibinfo{pages}{117--149}.
\bibitem[{Swendsen and Wang(1987)}]{PhysRevLett.58.86}
\bibinfo{author}{R.~H. Swendsen}, \bibinfo{author}{J.-S. Wang},
\newblock \bibinfo{title}{Nonuniversal critical dynamics in monte carlo simulations},
\newblock \bibinfo{journal}{Phys. Rev. Lett.} \bibinfo{volume}{58} (\bibinfo{year}{1987}) \bibinfo{pages}{86--88}.
\bibitem[{Wang and Landau(2001)}]{PhysRevLett.86.2050}
\bibinfo{author}{F.~Wang}, \bibinfo{author}{D.~P. Landau},
\newblock \bibinfo{title}{Efficient, multiple-range random walk algorithm to calculate the density of states},
\newblock \bibinfo{journal}{Phys. Rev. Lett.} \bibinfo{volume}{86} (\bibinfo{year}{2001}) \bibinfo{pages}{2050--2053}.
\bibitem[{Wolff(1989)}]{PhysRevLett.62.361}
\bibinfo{author}{U.~Wolff},
\newblock \bibinfo{title}{Collective monte carlo updating for spin systems},
\newblock \bibinfo{journal}{Phys. Rev. Lett.} \bibinfo{volume}{62} (\bibinfo{year}{1989}) \bibinfo{pages}{361--364}.
\bibitem[{Ferrenberg et~al.(2018)Ferrenberg, Xu, and Landau}]{PhysRevE.97.043301}
\bibinfo{author}{A.~M. Ferrenberg}, \bibinfo{author}{J.~Xu}, \bibinfo{author}{D.~P. Landau},
\newblock \bibinfo{title}{Pushing the limits of monte carlo simulations for the three-dimensional ising model},
\newblock \bibinfo{journal}{Phys. Rev. E} \bibinfo{volume}{97} (\bibinfo{year}{2018}) \bibinfo{pages}{043301}.
\bibitem[{Komura and Okabe(2012)}]{KOMURA20121155}
\bibinfo{author}{Y.~Komura}, \bibinfo{author}{Y.~Okabe},
\newblock \bibinfo{title}{Gpu-based swendsen–wang multi-cluster algorithm for the simulation of two-dimensional classical spin systems},
\newblock \bibinfo{journal}{Computer Physics Communications} \bibinfo{volume}{183} (\bibinfo{year}{2012}) \bibinfo{pages}{1155--1161}.

\end{thebibliography}

\section{Supplementary}
\subsection{Ising model}
\label{S:1}
Ising model is a fundamental model in statistical mechanics \cite{PhysRev.65.117}. The study of the Ising model not only helps elucidating a plethora of important physics phenomena, such as ferromagnetic phase transition, renormalization-group, and critical behaviour \cite{PhysRevB.29.4030}, but also leads to the invention of many widely-used algorithms, such as the Swendsen-Wang algorithm \cite{PhysRevLett.58.86}, Wang-Landau algorithm \cite{PhysRevLett.86.2050}, and Wolff cluster flipping algorithm \cite{PhysRevLett.62.361, PhysRevE.97.043301}. In the simplest form, the Ising Hamiltonian can be written as:
\begin{align}
    H(\boldsymbol{\sigma}) = - \sum_{\langle i, j\rangle} J\sigma_{i}\sigma_{j}, \label{Ising}
\end{align}
where $J$ is the interacting parameter and $\sigma_{i}$ represents the spin at the neighboring lattice site $i$, and $\boldsymbol{\sigma}$ denotes a spin configuration in the lattice. Note that the summation is over all sites and their nearest neighbors. For example, in a 2D square lattice, there are four nearest neighbors, and in a 3D square lattice, there are 6 nearest neighbors. 
It is also straightforward to generalize the above model to include longer-distance pair interactions within a predefined cutoff radius $r_c$ in a general lattice. Such Ising-like models can then be applied to study real materials, in which the spin variable $\sigma$ represents the different elements, and the generalized interaction parameters $J_{\sigma_i \sigma_j}$ can be interpreted as the chemical bonding between atoms of specific elements.

\subsection{Statistical mechanics}
\label{S:2}
In statistical mechanics, the probability for the occurrence of configuration $\boldsymbol{\sigma}_k$ is given by:
\begin{align}
    P(\boldsymbol{\sigma}_k) = \frac{e^{-\beta H(\boldsymbol{\sigma}_k)}}{Z}, \label{probability}
\end{align}
where the partition function is the sum of all the microstate probabilities:
\begin{align}
    Z = \sum_{\boldsymbol{\sigma}_k} e^{-\beta H(\boldsymbol{\sigma}_k)},
\end{align}
and $\beta = 1/(k_B T)$, where $k_B$ is the  Boltzmann constant, and $T$ is the temperature. After obtaining the partition function, physical observables $\langle \hat{O} \rangle$ such as the expected energy, magnetization, specific heat, and magnetic susceptibility can then be calculated accordingly from
\begin{align}
    \langle \hat{O} \rangle = \sum_k \frac{ O(\sigma_k)e^{-\beta H(\boldsymbol{\sigma}_k)}}{Z}.
\end{align}
\subsection{Monte Carlo Simulation Algorithms}
A sample following the probability distribution of Eq.~\ref{probability} can be generated via Markov chain Monte Carlo (MCMC) simulation. For such a purpose,  different schemes can be devised, and the Metropolis-Hastings algorithm is among the most commonly used ones. For the Ising model in Eq.~\ref{Ising}, the Metropolis-Hasting algorithm starts with proposing flipping the spins of each lattice site. This trial move gives rise to a change of the total energy by $\Delta E$, and the acceptance probability of this trial is given by:
\begin{align}
P=\begin{cases}
1, &\Delta E \leq 0\\
\exp(-\beta{\Delta E}), &\Delta E >0.
\end{cases}
\end{align}
A Monte Carlo step, or sweep, is defined as attempting the above trial exactly one time for each lattice site. A Monte Carlo simulation typically starts with a given number of warm-up steps. These steps are needed for the system to reach thermal equilibrium. After that a measurement of the configuration is made for each step, and the results can be recorded to calculate the physical observables.

The Monte Carlo simulation can be carried out at different conditions, the most common ones are the canonical ensembles, grand-canonical ensembles, and semi-grand canonical ensembles. In a canonical ensemble, the number of particles and the temperature are kept fixed. In the grand-canonical ensemble, the chemical potential is kept fixed, and the number of particles is determined by the equilibrium condition. The semi-grand canonical is a technique to fix the concentration of different elements by adjusting the value of the chemical potentials of each element. In this work, we focus on the canonical ensemble, which is relatively straightforward to implement. In a canonical Monte Carlo simulation, the first thing to note is that the flipping trial in the Ising model is replaced with a swapping trial, in which a random pair of sites is chosen randomly to switch the particles.

\subsection{Parallelization opportunities in Monte Carlo simulation}

{\color{black}
While there are many parallel opportunities in MC simulation, such as in temperature, and independent runs, the trial moves are generally challenging to parallelize due to the sequential nature of commonly used updating algorithms such as Metropolis, as pointed out in previous works \cite{sadigh2012scalable, OrtegaZamorano2013FPGAIsing, Liang2017GpuMonteCarlo}. } Nevertheless, there are still some parallelization opportunities that can be exploited. In general, the parallel opportunities in MC can be divided into three groups:
\begin{itemize}
    \item Temperature parallelism: the simulations at different temperatures are independent, therefore they are ``embarrassingly parallel". A closely related case is the replica exchange method, in which the configurations from different temperatures are constantly exchanged in the simulation process, with a policy satisfying detailed balance condition, in order to enhance the speed for reaching thermal equilibrium.
    
    \item Move parallelism: for two sites far-away from each other, the flipping or swapping move at one site won't affect the energy of the other, due to the short-range nature of the interactions. This property can therefore presents a parallelization opportunity. Similarly, it is also possible to split a huge lattice to smaller ones, as long as the ``ghost cells" at the boundary are taken care of.
    \item Sub-move parallelism: For calculating the energy change due to a single MC move, some fine-grain parallelization opportunities exist, depending on the form of the effective Hamiltonian. This is particularly true for the models based on neural networks, where a large number of matrix-multiplication operations present excellent opportunities for acceleration with highly optimized math libraries.
\end{itemize}
In this work, we mainly focus on move-parallelism, which is one of the bottleneck for large-scale MC simulations. For sub-move parallelism, we make use of the fact that, in the general case, all the local energies of sites within a LIZ need to be evaluated, and the calculations of these local energies are independent of each other.

{\color{black}
\subsection{Derivation of the scaling behavior}

\subsubsection{DFT}
The intrinsic scaling of DFT is well known to be approximately $O(N^3)$, where N typically refers to the number of electrons (and often used interchangeably with the number of atoms). This cubic scaling arises from matrix operations, such as diagonalization and matrix inversion, required to solve the Kohn–Sham equations.

\subsubsection{Traditional Monte Carlo simulation}
In standard MC implementations, one MC sweep involves sequentially updating all N sites. Without short-range approximations, each trial move requires computing the total energy, resulting in N energy evaluations per sweep. Combined with DFT’s $O(N^3)$ cost per energy evaluation, the total scaling becomes $O(N^4)$ for one MC sweep.

\subsubsection{Alternative MC strategies}
While the above applies to standard MC, alternative approaches also satisfy detailed balance and may demonstrate different scaling behaviors. For example, the Swendsen–Wang algorithm updates clusters of spins in Ising model \cite{KOMURA20121155}, and the Wang–Landau method samples random sites to estimate the density of states \cite{PhysRevE.97.043301}. While these methods are powerful for phase-transition studies, they are they are not general-purpose replacements for simpler MC algorithms like Metropolis, which are more straightforward and robust, particularly for simulating complex materials.

\subsubsection{Linear-Scaling DFT and near-sightedness}
The near-sightedness principle, introduced by Walter Kohn, states that quantum interactions are inherently short-ranged due to electron wave interference. This is the foundation of linear-scaling DFT methods such as LSMS, which restrict quantum interactions to a local interaction zone (LIZ). While this enables constant cost per local energy evaluation, the need to compute and sum energies across all atoms leads to an overall $O(N)$ cost. Combined with the N trials in one MC sweep, the total cost scales as $O(N^2)$

\subsubsection{MLPs}
Other than some GNN models, most modern MLPs are designed to be localized, such as HDNNP, MTP, and Allegro. Following the above analysis, the total cost scales as $O(N^2)$ due to a combination of $O(N)$ cost per total energy, and N trials in one MC sweep.

\subsubsection{SMC-X}
If long-range interactions are neglected (a common practice, but not always fully justified), then it becomes unnecessary to sum local energies over all atoms to compute the energy change for a Monte Carlo trial. This insight motivated our introduction of the local interaction zone (LIZ) concept and underlies the observed $O(N)$ scaling in Fig.~1(d). Again, note that similar idea has been previously employed in the SPMC algorithm \cite{sadigh2012scalable}, but has limits detailed in the introduction section of this work.
}

\end{spacing}
\end{document}